\documentclass[a4paper,11pt]{article}

\usepackage[utf8]{inputenc}
\usepackage[margin=1.1in]{geometry}
\usepackage{bm}
\usepackage{natbib,comment}
\usepackage{lscape}
\usepackage{float,enumitem}
\usepackage{amsmath,amssymb,amsthm,color,CJK,mathrsfs}
\usepackage[title]{appendix}
\usepackage{graphicx}
\usepackage{subcaption}
\usepackage{authblk}

\def\keywordfont{\fontsize{10}{11}\selectfont}
\newenvironment{keywords}{\par\addvspace{8pt}%
                          \keywordfont\noindent{\textbf{Keywords}}:\ \ignorespaces%
                         }{\par\addvspace{23pt}}

\newtheorem{algorithm}{Algorithm}

\newtheorem{proposition}{Proposition}

\newtheorem{definition}{Definition}

\DeclareMathOperator*{\argmax}{arg\,max}
\DeclareMathOperator*{\argmin}{arg\,min}
\newcommand{\E}{E}
\def\T{{ \mathrm{\scriptscriptstyle T} }}

\newcommand{\cov}{\text{cov}}
\newcommand{\var}{\text{var}}

\title{Variograms for spatial functional data with phase variation}

\author[1]{Xiaohan Guo}
\author[1]{Sebastian Kurtek}
\author[2]{Karthik Bharath}
\affil[1]{Department of Statistics, The Ohio State University}
\affil[2]{School of Mathematical Sciences, University of Nottingham}

\begin{document}
\date{}
\maketitle

\begin{abstract}
Spatial, amplitude and phase variations in spatial functional data are confounded. Conclusions from the popular functional trace variogram, which quantifies spatial variation, can be misleading when analysing misaligned functional data with phase variation. To remedy this, we describe a framework that extends amplitude-phase separation methods in functional data to the spatial setting, with a view towards performing clustering and spatial prediction. We propose a decomposition of the trace variogram into amplitude and phase components and quantify how spatial correlations between functional observations manifest in their respective amplitude and phase components. This enables us to generate separate amplitude and phase clustering methods for spatial functional data, and develop a novel spatial functional interpolant at unobserved locations based on combining separate amplitude and phase predictions. Through simulations and real data analyses, we found that the proposed methods result in more accurate predictions and more interpretable clustering results.  
\end{abstract}


\section{Introduction}
\subsection{Motivation}
In many disciplines, including environmental science, medicine, biology, geology and econometrics, it is increasingly common nowadays to observe functional data with complex spatial dependencies; such data are commonly referred to as spatial functional data \cite{delicado2010statistics}.
An archetypal example is the well-known Canadian weather data consisting of daily temperature recordings at 35 locations across Canada, described in detail in \citet{ramsay2004functional}. 
Data representing spatial functional data come in the form of traditional spatio-temporal data \citep{cressie2011statistics}. However, the functional data analysis framework allows one to directly capture temporal variation through its representation, thus enabling one to view data as discrete space-time realisations of a latent functional random field. 

From this perspective, spatial functional data analysis can be regarded as the extension of spatial statistical methods to functional data objects. While standard multivariate spatial statistics can be used once some form of dimension reduction of functional data objects has been carried out \citep{nerini2010cokriging}, the more popular approaches to model spatial correlations have been based on the notion of a metric-based trace-variogram \citep{giraldo2011ordinary}, which extends the standard variogram used in spatial statistics. Typically, the standard ${L}^2$ metric is used on a Hilbert space \citep{goulard1993geostatistical}. The trace-variogram, through a combination of the ${L}^2$ metric and the spatial distance, captures spatial dependencies between functional observations.

The trace variogram plays a central role in computation of spatially weighted discrepancies between, and weights that encode spatial correlatedness amongst, functional observations for clustering and prediction, respectively \citep{mateu2017advances}. A key assumption, implicit with the use of the ${L}^2$ distance in the trace-variogram, is that the temporal correspondence between functional observations is fixed. Thus, application of currently available $L^2$ metric-based trace-variogram methods to spatial functional data either assumes that the functions are perfectly aligned or treats phase variation as negligible noise. In reality, however, as with traditional functional data, it is frequently the case that the observed functions are out of phase: there is temporal misalignment of prominent geometric features of the functions, e.g., peaks and valleys. The adverse effects of disregarding phase variation while computing amplitude-related statistical summaries of functional data (e.g., functional mean and functional principal component analysis) using the ${L}^2$ distance are well-documented \citep{marron2015functional,srivastava2011registration}. The situation is exacerbated in the spatial setting since there are three sources of variation that are potentially confounded: amplitude, phase and spatial. For example, the comparison of average daily temperatures for two nearby cities in the Canadian temperature dataset 
should not only take into account the spatial dependency of seasonal high and low temperatures, but also temporal seasonal trends shared between the two cities.

While the issue of amplitude-phase separation has received considerable attention for traditional functional data (see  \citet{srivastava2016functional} and references therein), they are conspicuous in their absence within existing literature on spatial functional data comprising spatially correlated amplitude and phase components. Quantifying spatial variability with the trace-variogram thus requires its decomposition into separate amplitude and phase trace-variograms, based on a hitherto unavailable notion of spatially-informed amplitude and phase separation. In the presence of phase variation in the observed functions, such a decomposition will enable more interpretable clustering relating to amplitude and phase components, and will result in better prediction of functions at unobserved locations. 

\subsection{Contributions and related work}

To the best of our knowledge, this is the first attempt to model spatial functional data with phase variation, via separate definitions of the trace-variogram for amplitude and phase components. The central challenge lies in synthesizing spatial information with the fundamental asymmetry between the absolute and relative notions of phase and amplitude of a function, respectively, in order to develop a practically viable decomposition of the trace-variogram for clustering and prediction. Accordingly, our contributions are as follows.  
\begin{enumerate}[leftmargin=*]
\itemsep 0em
    \item We define an amplitude trace-variogram on the spatial domain, and in order to account for the relative nature of phase, we define a conditional phase trace-variogram on an augmented domain comprising shape of the observed functions as a covariate.
    \item We propose an algorithm to compute a spatially-weighted mean, which enables joint alignment of functions and computation of estimators of the amplitude and phase trace-variograms.
    \item Based on the variograms, we propose (i) linear unbiased estimators for spatial prediction or kriging of amplitude and phase (and combine them to form the final prediction), and (ii) a method for clustering spatial functional data into amplitude and phase clusters. Our framework treats spatial phase variation as a key feature of spatial functional data rather than noise.
\end{enumerate}

Adaptation of multivariate spatial data methods to functional clustering, following dimension reduction, was done in 
\citet{giraldo2012hierarchical} and \citet{haggarty2015spatially}. \citet{romano2010clustering,romano2017spatial} extended the classical dynamic clustering approach in geostatistics to spatial functional data by employing the trace-variogram. 
On the other hand, \citet{secchi2013bagging} introduced Bagging Voronoi-classifiers for clustering spatial functional data. This method was further improved by \citet{abramowicz2017clustering} by combining it with $k$-means registration \citep{sangalli2010k}. 

Kriging or spatial prediction is based on borrowing information from nearby objects to construct predictions at new spatial locations; the contribution to the predictor from each function depends on the strength of spatial correlation. \citet{giraldo2011ordinary} used the trace-variogram for ordinary kriging of functional observations, which inspired related approaches. Chief amongst these are universal kriging methods \citep{caballero2013universal,menafoglio2013universal,reyes2015residual,menafoglio2016kriging} wherein observed functions are pre-processed to better manage deviations from the stationarity assumption. \citet{menafoglio2018kriging} generalized kriging of functional data to data on a Riemannian manifold.

\color{black}

\section{Amplitude-Phase Separation}\label{sec:apsep}
\subsection{Relevant function spaces and distances}
We build on the metric-based elastic functional data analysis framework \citep{srivastava2011registration,srivastava2016functional} for amplitude-phase separation. 
Without loss of generality, we consider the representation space of functional data objects to be $\mathcal F=\{f:[0,1]\to  R\mid f \text{ is absolutely continuous}\}$. The group of warping functions representing phase is $\Gamma=\{\gamma:[0,1]\to[0,1]\mid\gamma(0)=0,\gamma(1)=1,\dot\gamma>0\}$ ($\dot\gamma$ is the time derivative of $\gamma$). For any $f\in \mathcal F$, $\gamma\in \Gamma$, the warping of $f$ by $\gamma$ is given by the group action of composition, $f\circ\gamma$. The group-theoretic formulation of phase enables a definition of the amplitude of a function $f$ as the equivalence class $[f]=\{f\circ \gamma\mid\gamma\in \Gamma\} \subset \mathcal F$, known as its orbit under the action of $\Gamma$; thus, $f\circ \gamma \in [f]$ has the same amplitude as $f$ for each $\gamma \in \Gamma$. The amplitude space then is the quotient $\mathcal F/\Gamma=\{[f]\mid f\in\mathcal F\}$.  

Separating amplitude and phase requires a metric on the amplitude space $\mathcal F/\Gamma$. A convenient way to define one is through a metric $d$ on $\mathcal{F}$ that is invariant to simultaneous warpings: for every $\gamma\in\Gamma,\  d(f_1,f_2)=d(f_1\circ\gamma,f_2\circ\gamma)$. It is well-known that the standard $L^2$ metric fails to be invariant; \citet{srivastava2011registration} thus proposed the isometric Fisher-Rao metric. Unfortunately, this metric is difficult to use in practice. However, the \emph{square-root slope transform} remarkably reduces the complicated Fisher-Rao metric on $\mathcal F$ to the standard $L^2$ metric on the transformed space. The transform maps $f \mapsto Q(f)=q=\text{sgn}(\dot f)|\dot f|^{1/2}$ ($\dot f$ is the time derivative of $f$). Given $f(0)$, $Q$ is bijective with inverse $Q^{-1}(q,f(0))(t)=f(t)=f(0)+\int_0^tq(u)|q(u)|du$. 

The transformed space $Q(\mathcal F)$ is a subset of $L^2$, and is denoted by $\mathcal Q$. Under $Q$, the Fisher-Rao metric on $\mathcal F$ maps to the standard $L^2$ metric on $\mathcal Q$, and thus analysis of square-root slope transformed functional observations can be carried out using standard Hilbert space machinery. Warping of $f \in \mathcal F$ by $\gamma$ induces the warping action $(q,\gamma)=(q\circ\gamma){\dot\gamma}^{1/2}$ on $\mathcal Q$ with corresponding orbit or amplitude $[q]:=\{(q,\gamma)|\gamma \in \Gamma \}$ and amplitude space 
$\mathcal Q/\Gamma=\{[q]\mid q\in\mathcal Q\}$. 
\begin{definition}[Amplitude and Shape distance]\label{ampdist}
The amplitude distance between $q_1,q_2\in\mathcal Q$ is defined as
$d_{a}(q_1,q_2)= \inf_{\gamma\in\Gamma}\ \|q_1-(q_2,\gamma)\|.$ 
The shape distance between $q_1, q_2 \in \mathcal Q$ is defined as
$d_{sh}(q_1,q_2)=d_{a}(q_1/\|q_1\|,q_2/\|q_2\|)$.
\end{definition}

Amplitude and phase separation through registration or alignment of $f_2$ to $f_1$ (or vice versa) is formulated as the determination of the relative phase obtained by solving
\begin{equation}\label{align}
\gamma^*=\underset{\gamma\in\Gamma}\argmin \ \|q_1-(q_2,\gamma)\|, 
\end{equation}
typically using the dynamic programming algorithm,
where $q_1,q_2$ are the square-root transformed $f_1,f_2$. The optimal alignment of $f_2$ with respect to $f_1$ is given by $f_2\circ\gamma^*$, and \eqref{align} specifies the pairwise alignment problem. Operationally, we will thus refer to $f \circ \gamma^*_i$ and $(q,\gamma^*_i)$ as the amplitude of $f$ and $q$ respectively. Joint registration of $f_1,\dots,f_n$, with respect to a known template, is carried out by pairwise alignment of each function in the sample with respect to the template. In the absence of a template, the Karcher mean is used \citep{srivastava2011registration}. 

Alignment of $f_2$ to $f_1$ using $q_1$ and $q_2$ allows us to compute their relative phase distance. For this, we consider the square-root transform $\psi$ of $\gamma$: $\gamma \mapsto Q(\gamma)=\psi=\dot\gamma^{1/2}$. Since $\int_0^1 \psi^2(t)dt=1$, the square-root transformed warping group $Q(\Gamma)=\Psi$ is the positive orthant of the unit sphere in ${L}^2[0,1]$, enabling us to consider intrinsic and extrinsic relative phase distances. 
\begin{definition}[Phase distances]\label{phasedist}
If $\psi^*=\dot\gamma^{1/2}$ is the relative phase between $q_1,q_2 \in \mathcal Q$, then their intrinsic relative phase distance is $$d^{int}_{p}(q_1,q_2)=\cos^{-1}\left(\int_0^1 \psi^*(t)\psi_{id}(t)dt\right),$$
where $\psi_{id}(t)=1$ is the square-root transformed identity warping function $\gamma_{id}(t)=t$. However, the extrinsic phase distance between $\psi_1,\ \psi_2\in\Psi$ is $\| \psi_1-\psi_2\|$.
\end{definition}
Due to the nonlinear nature of time warping, the usual ${L}^2$ distance between $q_1,q_2 \in \mathcal Q$ does not decompose exactly into the respective amplitude and phase distances in Definitions \ref{ampdist} and \ref{phasedist}. The elastic framework, however, enables us to extract pure amplitude and phase components, and disentangle them from spatial variation in spatial functional data. 
\subsection{Setup and notation}
The setting throughout this paper is that of geostatistical dense functional data \citep{wang2016functional}, wherein a function at each spatial location is assumed to have been observed on a fine partition of $[0,1]$. We assume a square-integrable functional random field $\{f_s: s \in \mathcal{D}\}$ on a spatial domain $\mathcal{D} \subseteq {R}^2$; see, e.g., \citet{menafoglio2013universal} for formal definitions. Associated with $f_s$ is its square-root slope transformed version $\{q_s, s \in \mathcal D\}$ such that $s \mapsto q_s \in \mathcal{Q}$.

Observed functional data $f_{s_i}, s_i \in \mathcal D\ (i=1,\ldots,n)$ will be transformed using the square-root slope transform to obtain $q_{s_i}$, and methodology will be entirely developed using the $q_{s_i}$. Henceforth, the subscript $i$ as an index is short for the spatial location $s_i$ (e.g., $q_i,\ \gamma_i$); the subscript $s$ will only be used with a functional random field (e.g., $q_s$). The ${L}^2$ norm on the function spaces $\mathcal Q$ and $\Psi$ will be denoted by $\|\cdot\|$, while $\|\cdot\|_2$ will denote the Euclidean norm on $\mathcal D$.  

\section{Amplitude-phase separation of trace variogram}
\label{sec:ampphvar}
\begin{figure}
    \centering
    \includegraphics[scale=0.22]{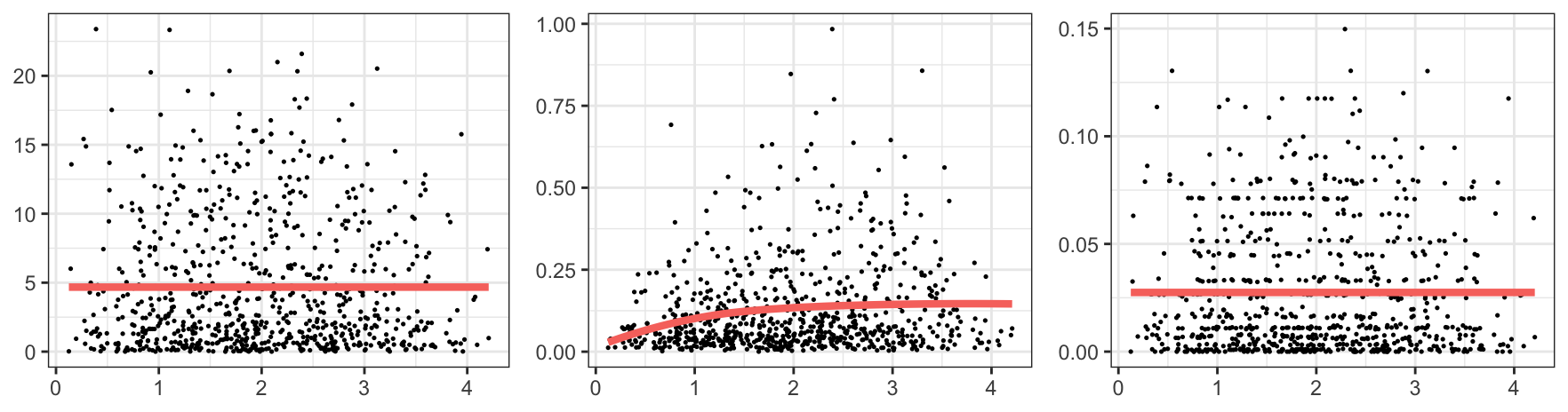}
    \caption{Decomposition of the trace-variogram (left) into amplitude (middle) and phase (right) components for simulated functional data with spatially correlated amplitudes and nearly independent phases. The trace-variogram fails to detect spatial correlations. Estimates of the trace-variograms (red curves) are obtained by fitting a valid parametric model to an empirical estimator (see Section \ref{sec:fitting}). }
    \label{fig:separation}
\end{figure}

The trace-variogram of a functional random field $\{f_s, s \in \mathcal D\}$ is defined as $V(s,s')=0.5E(\|f_s-f_{s'}\|^2)$. If $E(f_s(t))=\mu(t)$ and thus constant in space, and if for every $t,t'\in[0,1]$, $s,s' \in \mathcal D$, $\text{cov}(f_s(t),f_{s'}(t'))$ is a function of the spatial distance $h=\|s-s'\|$ only, the random field is said to be \emph{second-order stationary and isotropic}, and the trace-variogram reduces to $V(h)$  \citep{giraldo2011ordinary}. Its definition based on the pointwise ${L}^2$ distance thus implicitly assumes that $f_s$ and $f_{s'}$ are registered with zero phase variation. However, such an assumption is unrealistic in most real data settings as true spatial function variability is often confounded with amplitude and phase variation. 
The left panel of Figure \ref{fig:separation} demonstrates this issue on 
simulated functions from a second-order stationary and isotropic functional random field, wherein the spatial dependency arises primarily through the amplitude component and not through the phase.
Failure to disentangle amplitude and phase variations leads to a trace-variogram $V(h)$ (left panel) that suggests negligible spatial dependency between the functions. On the other hand, decomposing the trace variogram into amplitude and phase components (Definition \ref{eq:amp_vargm}) captures the correct form of spatial correlatedness (middle and right panels in Figure \ref{fig:separation}). In the absence of consistent spatial dependency patterns, it is necessary to first decouple the amplitude and phase components in spatial functional data prior to quantifying spatial correlation.  
 
In order to decompose the trace-variogram, consider the model
\begin{equation}
\label{model2}
q_{i}(t)=(\mu_{q} +e_{i},\gamma_i^{-1})(t)=\left\{(\mu_{q} +e_{i}) \circ \gamma^{-1}_{i}\right\}(t)\left\{\dot{\gamma}^{-1}_{i}(t)\right\}^{1/2},\ t \in [0,1]\thinspace,
\end{equation}
for the observed data $(i=1,\ldots,n)$ with unobserved $\mu_q$ and $\gamma_i$.  Equivalently, $(q_i,\gamma_i)(t)=\mu_q(t)+e_i(t)$. The process $\{(q_s,\gamma_s), s \in \mathcal D\}$ is the \emph{amplitude random field}, and $\{\gamma_s, s \in \mathcal D\}$ the corresponding \emph{phase random field}, associated with the functional random field $\{q_s, s \in \mathcal D\}$. The $\mu_q \in \mathcal Q$ is a deterministic mean amplitude, constant in space, and $e_s$ and $\gamma_s$ are mutually independent functional random fields assuming values in $\mathcal Q$ and $\Gamma$, respectively, with $\E(e_s)\equiv 0$ and $\E (\gamma_s(t))=\gamma(t)$ for all $s \in \mathcal D$. The random field $\gamma_s$ depends on $q_s$ only through the mean $\mu_q$. 

Under the model, the amplitude of $q_i$ is $(q_i,\gamma_i)$ and $\mu_q$ is thus the common mean amplitude. If $\mu_q$ is known, then after obtaining estimates $\hat \gamma_i$ through an alignment procedure, $(q_i,\hat \gamma_i)$ can be analyzed using the trace-variogram.
An unknown $\mu_q$, however, cannot be consistently estimated since only its orbit $[\mu_q]$ is identifiable, with the exception being when the model comprises only scale variation \citep{KS, CP}. We thus use $\mu_q$ merely as a conceptual device to elucidate decomposition of the trace-variogram, and do not estimate it. The $q_i$ are aligned instead using a local procedure outlined in Section \ref{kriging_sec}.

Conditional on $\gamma_s$, it may be reasonable to assume that the amplitude random field $(q_s,\gamma_s)$ is second-order stationary and isotropic on $\mathcal D$. This is tantamount to assuming that nonstationarity manifests in $q_s$ purely through time warping. On the contrary, since $\gamma_s$ and $\gamma_{s'}$ are correlated not only spatially, but also through how similar the shapes of $q_s$ and $q_{s'}$, stationarity of $\gamma_s$ is almost never satisfied; in other words, spatially proximate functions with similar amplitudes or shapes are more likely to exhibit similar phase components. For example, patterns of seasonal temperature highs and lows tends to be similar in phase for nearby locations. 

Nonstationarity of $\gamma_s$ on $\mathcal D$ can be addressed by introducing covariate information, as done in \citet{schmidt2011considering}. The relevant covariate is the amplitude or perhaps shape of $q_s$. Accordingly, we consider the englarged domain $\mathcal D'=\mathcal D \times \mathcal Q$, wherein `spatial lag' is defined for $y_1,y_2 \in \mathcal D'$ as $y_1-y_2=(s_1-s_2,q_1-q_2)$, and
equip it with the squared distance metric 
\begin{equation}
\|y_1-y_2\|^2_\omega=\|s_1- s_2 \|^2_2+\omega d^2_{sh}(q_1,q_2), 
\end{equation}
with tuning parameter $\omega \geq 0$. The metric is motivated by the observation that $\gamma_{s}$ and $\gamma_{s'}$ will be similar if $q_s$ and $q_{s'}$ have similar amplitudes; in other words, high correlation between $\gamma_{s}$ and $\gamma_{s'}$ is driven by a small phase distance $d_p^{int}(q_s,q_{s'})$, which occurs when $d_{sh}(q_s,q_{s'})$ is small through $d_{sh}(q_s,q_{s'}) \leq d_{sh}(q_s,\mu_q)+d_{sh}(q_{s'},\mu_q)$. The value of $\omega$ adjusts the effect of this phenomenon with the spatial proximity between  $\gamma_{s}$ and $\gamma_{s'}$. However, since $q_s$ is a random field, the set $\mathcal D'$ as a domain only makes sense when conditioned on $\{q_s,\ s \in \mathcal D\}$. Thus, when conditioned on $q_s$, it may be reasonable to assume that $\gamma_y$, or its square-root slope transformed version, is second-order stationary and isotropic on $\mathcal D'$. 

\begin{definition}
\label{amp_vargm}
Suppose that the random fields $\{(q_s,\gamma_s), s \in \mathcal D\}$ and $\{\psi_y=\dot \gamma^{1/2}_y, y \in \mathcal D'\}$ are second-order stationary and isotropic. 
\begin{enumerate}
    \item 
The amplitude trace-variogram is defined as 
\begin{align}\label{eq:amp_vargm}
\|s-s'\|_2=h \mapsto V_a(h)=\frac{1}{2}\E\left(\|(q_s,\gamma_s)- (q_{s'},\gamma_{s'})\|^2\right).
\end{align}
\item 
The phase trace-variogram is defined as 
\begin{align}
\|y-y'\|_\omega=h \mapsto V_p(h)=\frac{1}{2}\E\left(\|\psi_y-\psi_{y'}\|^2\right).
\end{align}
\end{enumerate}
\end{definition}

\section{Amplitude-phase Kriging}\label{kriging_sec}
\subsection{Amplitude kriging with spatially weighted mean amplitude}\label{amp_kriging_sec}

\citet{giraldo2011ordinary} developed a linear unbiased estimator that extends ordinary kriging or spatial prediction to the functional setting by minimizing the $L^2$ prediction error. In the presence of phase variation, the ${L}^2$-based linear estimator can be biased, since functional features such as local extrema can be misaligned. Examples in Figure \ref{fig:simu_kriging_map2_bsp} illustrates this phenomenon.

Given pairs $\{(s_i,q_i) \mid s_i\in \mathcal D\}\  (i=1,\dots,n)$, the goal is to predict an unobserved function $q_0$ at a new location $s_0 \in \mathcal D$ comprising amplitude $(q_0,\gamma_0)$ and phase $\gamma_0$. To address possible misalignment of $q_i$, we consider a three-stage kriging procedure: (1) predict the amplitude component; (2) predict the phase component related to the predicted amplitude; and (3) combine the two to obtain the kriging estimate.
  
We propose a linear amplitude kriging estimator using an iterative procedure (Algorithm 1) which in each iterate proposes a local data-driven template for aligning the $q_i$, computes the linear estimator using the aligned $q_i$, and uses the estimator in the current iterate as the template for the next. Output from the algorithm is thus
a \emph{spatially-weighted amplitude estimator that serves the dual purpose of acting as a local template for alignment and as an estimate of the amplitude component $(q_0,\gamma_0)$}. The procedure thus circumvents the issues associated with estimating $\mu_q$ in model \eqref{model2} described in Section \ref{sec:ampphvar} for use as a template for alignment. 

In anticipation of the computing required within each iterate of Algorithm 1 below, we define the amplitude estimator with respect to some fixed template, say $q$. With $\hat \gamma_i$ as the estimated phases following alignment of $q_i$ with $q$, let $(q_i,\hat\gamma_i)$ be the estimates of amplitude of $q_i$. Let $\Delta_n:=\{(x_1,\ldots,x_n)^\T \in R^n|\sum_{i=1}^n x_i=1\}$. Our linear estimator of the amplitude $(q_0,\gamma_0)$ is
\begin{equation}\label{eq:amp_est}
\tilde q_0(t)=\sum\limits_{i=1}^n \eta_i(q_i,\hat\gamma_i)(t),
\end{equation}
where the coefficient vector $\eta =(\eta_1,...,\eta_n)^\T \in \Delta_{n}$  is implicitly defined as the minimizer of the expected amplitude prediction error functional
\begin{equation} \label{eq:amp_pred_error}
\eta \mapsto \E\left(\|\tilde q_0-(q_0,\gamma_0)\|^2\right).
\end{equation}

\begin{proposition}
\label{prop:pred_error}
Given a template $q$, suppose $E((q_i,\hat\gamma_i))=q$ for each $i=1,\ldots,n$. 
Then the $\eta \in \Delta_n$ that minimizes \eqref{eq:amp_pred_error} also minimizes $\eta \mapsto  \eta^\T   \mathcal{V}_a    \eta$,
where the $n \times n$ matrix $\mathcal{V}_a$ contains as its elements   $V_a(h_{0j})+V_a(h_{i0})-V_a(h_{ij})$ with $h_{ij}=\| s_i- s_j\|$ $(i=1,\ldots,n;\  j=1,\ldots,n)$. The amplitude predictor in \eqref{eq:amp_est} thus depends only on the amplitude trace-variogram $V_a(h)$. 
\end{proposition}

Computing $\eta$ using Proposition \ref{prop:pred_error} requires knowledge of the amplitude varigram $V_a$. The plug-in nonparametric estimate of $V_a$ is given by
\begin{align}
\widehat V_a(h)=\frac{1}{2|N(h)|}\sum \limits_{i,j\in N(h)} \|(q_i,\hat\gamma_i)-(q_j,\hat\gamma_j)\|^2,
\end{align}
where $N(h)=\{(  s_i,  s_j)\mid \|  s_i-  s_j\|=h\}$. For irregularly spaced data, $N(h)$ can be modified to $\{(  s_i,  s_j):\|  s_i-  s_j\|\in (h-\epsilon,h+\epsilon)\}$ for a small $\epsilon >0$. The iterative algorithm to compute the amplitude kriging estimate is as follows. 
\begin{algorithm}{(Amplitude kriging estimate)}
\label{algorithm1}

Input: $q_1,\ldots,q_n$; Output: Amplitude kriging estimate $\tilde q_0$.

Step 1. Set $k=0$ and initialize the template $\hat q_0^{(0)}$  with the $q_i\ (i=1,\ldots,n)$ 

   \qquad   \qquad  
closest to $s_0 \in \mathcal{D}$. 
 
 Step 2: Repeat:
 
  \qquad \qquad
        Align each $q_i$ to $\hat q_0^{(k)}$ to get $(q_i,\gamma_i^{(k)})$ using procedure in \eqref{align};
 
   \qquad   \qquad  
        Compute $\widehat V_a(h)$ using $\{(q_i,\gamma_i^{(k)})\}$ and 
        $\tilde q_0^{(k)}=\sum_{i=1}^n \eta_i (q_i,\gamma_i^{(k)})$ 
    
    \qquad   \qquad  
        via Proposition \ref{prop:pred_error};
  
    \qquad    \qquad  
        Set $\hat q_0^{(k+1)}=\tilde q_0^{(k)}$;
        
    \qquad \ \ \ \
    Until $\|\hat q_0^{(k+1)}-\hat q_0^{(k)}\|<\epsilon$, for some small tolerance $\epsilon$.
\end{algorithm}

\noindent Within each iteration $k$, the template $\hat q^{(k)}_0$ is fixed, and acts as the given template $q$ used in Proposition \ref{prop:pred_error}. Spatial information is incorporated in Step 2 (third line) through the use of $\hat V_a(h)$. As with any estimator of the mean amplitude $\mu_q$ in model \eqref{model2}, consistency of $\tilde q_0$ as an estimator of amplitude $(q_0,\gamma_0)$ can be established for the restricted one-dimensional amplitude model.


\begin{proposition} 
Consider observations $q_1,\ldots,q_n$ from the simplified model $(q_s,\gamma_s)(t)=c_s \mu_q,$, where $c_s$ is a scalar, positive, isotropic random field on $\mathcal D$. Suppose the variogram of $c_s$ is known and continuous in a neighborhood of $0$, and $s_0$ is a limit point of $\{s_1,...,  s_n\}$ as $n\to\infty$. Then, the estimator $\tilde q_0$ of $q_0$ under the model obtained using Algorithm \ref{algorithm1} converges in $L^2$ to an element in the orbit $[q_0]$ as $n\to\infty$.
\end{proposition}


\subsection{Phase kriging and combined prediction} \label{sec:phase_kriging}
In amplitude kriging, phase variability is removed by aligning all functions with respect to the estimated template, which results in improved prediction of the  \emph{shape} and \emph{magnitude} of a function. However, $\tilde q_0$ is the prediction of $(q_0,\gamma_0)$ rather than $q_0$. Thus, to obtain the final prediction of $q_0$, we construct an estimator of $\gamma_0$ via phase kriging, using \emph{the estimated phase $\hat\gamma_i\in\Gamma$, computed by aligning $q_i$ to the amplitude kriging estimate $\tilde q_0$} with corresponding square-root slope transforms $\hat\psi_i\in\Psi$ $(i=1,\ldots,n)$. 

We want to predict $\psi_0$ on $\Psi$, which is nonlinear, using the relative phases $\hat\psi_1,...,\hat\psi_n$. We deal with the nonlinearity of $\Psi$ by considering the positive  extension $\Psi'=\{\psi'=a\psi\mid a\in R^+,\psi \in\Psi\}$ of $\Psi$. Compatible with the linearity of the amplitude kriging estimate $\tilde q_0$, we compute the corresponding linear phase kriging estimate in $\Psi'$ and then project it back to $\Psi$. The projection $\Pi: \Psi'\to\Psi$ is defined as $\Pi(x)={\argmin}_{\psi\in\Psi}{\|\psi-x\|}=x/{\|x\|}$. Thus 
 $\Pi(\tilde\psi_0)={\tilde\psi_0}/{\|\tilde\psi_0\|}$ is the phase kriging estimator of $\psi_0$ based on a linear estimator $\tilde \psi_0 \in \Psi'$. 

Let $\Delta^+_n:=\{(x_1,\ldots,x_n)^T \in R^n|x_i >0,\ \sum_{i=1}^n x_i=1\}$. With the conditional random field $\psi_y$ on $\mathcal D'$ equipped with distance $\|\cdot\|_\omega$, the linear estimate of $\psi_0$ in $\Psi'$ is defined as
\begin{equation}\label{phase_pred}
\tilde \psi_0(t)=\sum\limits_{i=1}^n\zeta_i\hat\psi_i(t)\thinspace,
\end{equation}
where $\zeta=(\zeta_1,\ldots,\zeta_n)^T \in \Delta^+_n$ minimizes the phase prediction error functional, defined as in \eqref{eq:amp_pred_error} using $\psi_0$ and $\tilde \psi_0$. Positivity of $\zeta_i$ is required to ensure that the resulting warping functions are strictly increasing. As in Proposition \ref{prop:pred_error} for amplitude kriging:
\begin{proposition}
\label{prop:phase_ pred_error}
The vector $\zeta \in \Delta^+_n$ can be obtained by minimizing $\zeta \mapsto  \zeta^\T   \mathcal{V}_p    \zeta$,
where the $n \times n$ matrix $\mathcal{V}_p$ contains as its elements   $V_p(h_{0j})+V_p(h_{i0})-V_p(h_{ij})$ with $h_{ij}=\| y_i- y_j\|_\omega$ $(i=1,\ldots,n;\  j=1,\ldots,n)$. 
\end{proposition}
The plug-in nonparametric estimator of the phase trace-variogram is
\begin{align}\label{eq:emp_phase_vargm}
\widehat V_p(h)= \frac{1}{2|N(h)|}\sum\limits_{i,j\in N(h)} \|\hat\psi_i-\hat\psi_j\|^2,\quad N(h)=\{(i,j)|g_{\omega,ij}=h\}.
\end{align}

Although $\hat V_a$ and $\hat V_p$ are related through the alignment of $\{q_i\}$, they can exhibit different patterns as dictated by the structure of the spatial dependence between the $\{q_i\}$

The predicted amplitude and phase kriging estimates $\tilde q_0$ and $\tilde \psi_0$ include all information about the magnitude, shape and temporal characteristics of the final prediction, but not the translation, which is lost due to the square-root slope transformation. To account for this, we use the starting points $f_i(0)\  (i=1,\ldots,n)$ of the observed functions and apply ordinary kriging \citep{cressie2011statistics} to obtain a translation prediction estimate $\tilde T_0$ of the unknown function $f_0$. 
\begin{figure}
\centering
\includegraphics[scale=0.15]{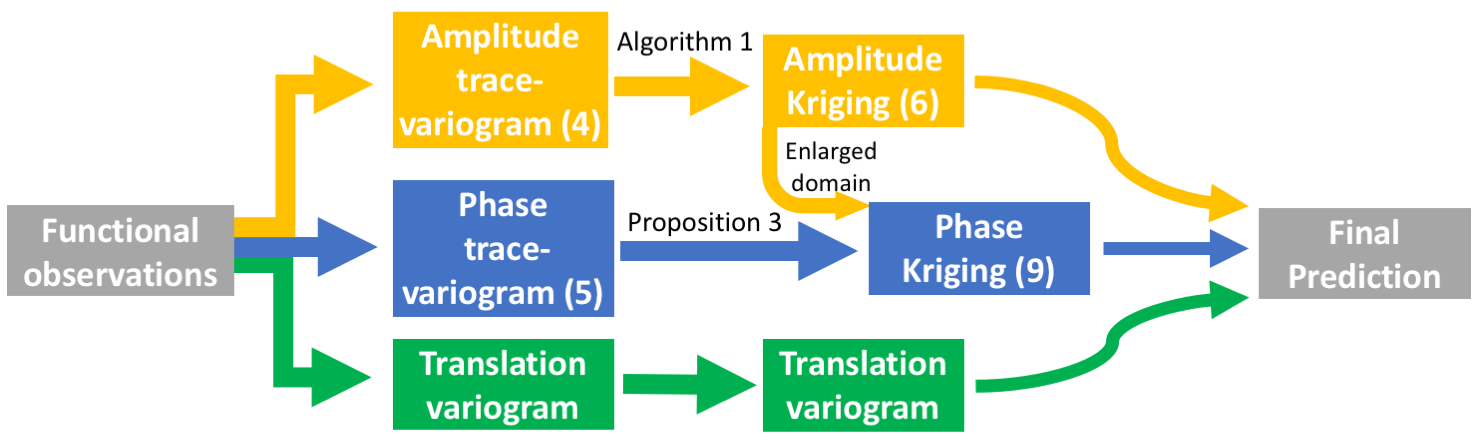}
\caption{Pipeline for amplitude-phase kriging procedure.}
\label{fig:krigingproc}
\end{figure}

Recall the inverse of the square-root slope transformation $Q^{-1}: (R \times \mathcal Q) \to \mathcal F$ from Section \ref{sec:apsep}. The final kriging estimate combines the three estimates of amplitude, phase and translation as follows.
First, we combine the amplitude and phase predictions using $ q_0^*=(\tilde q_0,\tilde \gamma^{-1}_0)$, where $\tilde \gamma_0(t)=\int_0^t \Pi(\tilde \psi_0(u))^2du$ is the phase prediction. The combined kriging estimate of $f_0$ at site $s_0$ then is $f_0^* = Q^{-1}(q_0^*,\tilde T_0)$, where $\tilde T_0$ is the predicted starting point. The full pipeline of the proposed kriging approach is shown in Figure \ref{fig:krigingproc}.

\section{Amplitude-phase Clustering} \label{sec:cluster}
Amplitude and phase distances arising from the amplitude-phase separation enable separate distance-based amplitude and phase clustering of functional data. Spatially informed adaptations can now be defined through the use of dissimilarity measures by combining the amplitude (phase) distance and amplitude (phase) trace-variogram. This can lead to more interpretable clusters. 
For example, in the Canadian weather data we note that daily average temperatures at sites with similar extreme temperatures (similar amplitude) need not experience similar seasonal trends. Thus, one would reasonably expect different clustering results corresponding to the two components. 

While several distance-based clustering approaches can be used, we consider a hierarchical clustering based on spatially weighted dissimilarity matrices \citep{giraldo2012hierarchical}. The amplitude dissimilarity matrix is defined using the distance
$d_{A,ij}=d_{a}(q_i,q_j)\times V_a(\|  s_i-  s_j\|)$, and the phase dissimilarity matrix is defined using the distance
$d_{P,ij}=d^{int}_{p}(q_i,q_j)\times V_p(\|y_{i}- y_{j}\|_\omega)$,
based on the enlarged spatial domain $\mathcal D'$. 

The dissimilarity matrices measure the discrepancy in amplitude and phase for each pair of functions. Thus, in this case, it is not necessary to choose a common template for all of the functions for alignment. Instead, we simply choose one of the functions in each pair as a template to compute the amplitude and phase distance between them. Then, the amplitude and phase trace-variograms in Definition \ref{amp_vargm} can be simplified, and with corresponding estimators 
\begin{equation*}
\widehat V_a(h)=\frac{1}{2|N_a(h)|}\sum \limits_{i,j\in N_a(h)} d_{a}(q_i,q_j)^2,\ \ \ \  \widehat V_p(h)= \frac{1}{2|N_p(h)|}\sum\limits_{i,j\in N_p(h)}d^{int}_{p}(q_i,q_j)^2,
\end{equation*}
where $ N_a(h)=\{(i,j)|h=\|s_i-s_j\|\}$ and $ N_p(h)=\{(i,j)|h=\|y_i-y_j\|_\omega\}$. Finally, the amplitude and phase dissimilarity matrices can be input in separate hierarchical clustering using the methods of \citet{everitt2001m}.

\section{Simulations}
\subsection{Fitting amplitude and phase trace-variograms}
\label{sec:fitting}
As in classical geostatistics, the variogram estimators $\hat V_a$ and $\hat V_p$ can fail to be conditionally negative definite, and thus it becomes necessary to fit a valid model. In simulations and real data examples, we fit Mat\'ern models with the smoothness parameter fixed to $0.5$ \citep{cressie2011statistics}. The tuning parameter $\omega$ in the enlarged domain $\mathcal D'$ for the phase trace-variogram is chosen as the one maximizing the goodness of fit. To increase the robustness of phase kriging in the case where large shape variation is present in the spatial functional data, we use a penalized alignment method (see Appendix \ref{sec:s2.1}) to estimate the relative phase functions. The tuning parameter for the penalty in that optimization problem is determined by cross-validation. 

\subsection{Kriging}  
We compare the proposed approach to ordinary kriging \citep{giraldo2011ordinary} using different types of simulated spatial functional data. We fix the spatial locations to equally-spaced sites on a $5\times 5$ grid with $x,y$ coordinates taking the values $(-2,-1,0,1,2)$. 
The simulated functional data $f_i\ (i=1,\ldots,25)$ are generated using the model
$f_{i}(t)=\{\sum_{j=1}^K(a_{i,j}B_j+e_i)\circ\gamma_i\}(t)$, where, for each $j$, the coefficient vector $[a_{1,j},\dots,a_{25,j}]$ follows a multivariate normal distribution with some mean and the Matern covariance $C_{Mat}(\cdot,\cdot; \sigma_a^2,0.5,\ell_1)$; here, $\sigma_a^2$ is the scale parameter, $\ell_1$ is the range, and the smoothing parameter is fixed to $0.5$. This imposes spatial correlation in the amplitude component of the simulated data. Holding $i$ fixed, the coefficients for the basis $B_j,\ j=1,\dots,k$ are assumed to be independent. We use two different choices of basis functions: (1) B-spline: set $K=10$ and $\{B_j\}_{j=1}^K$ to be cubic B-splines on $[0,1]$ with the mean of the coefficient vector for each $i$ equal to $(1,2,3,4,5,5,4,3,3,2,1)^\T$; (2) Bimodal: set $K=1$ and $B_1(t)=-\cos( 2\pi t)$ on $[-1,1]$, with the mean of the coefficients $a_{1,1},\dots,a_{25,1}$ set to 5. The phase components $\gamma_i,\ i=1,\dots,25$ are distribution functions of $Beta(1,e^{b_i})$ with $\{b_1,...,b_{25}\}$ generated from the correlated uniform distribution on $[-B,B]$ by transforming a random sample from the multivariate normal distribution with covariance $C_{Mat}(\cdot,\cdot;1,0.5,\ell_2)$. This, in turn, generates spatially correlated phase functions. In this phase model, the parameter $B$ determines the magnitude of phase variation and $\ell_2$ controls the range of spatial dependency. We let $\ell_1 =\ell_2=2\surd2$. Each error term $e_i$ is generated from a white noise process with variance $0.25$.


We perform leave-one-out cross-validation with ${f}^{[-i]*}$ denoting the prediction of $f_i$ using all observations except the $i$th one. To measure the accuracy of predictions, we compute the following five error metrics:
\begin{itemize}
\item Amplitude least squares: $E1=n^{-1}\sum_{i=1}^n \|f_*^{[-i]}-f_i\|^2
$, where $f_*^{[-i]}$ is ${f}^{[-i]*}$ after optimal alignment to $f_i$; 

\item Amplitude Sobolev least squares:
$E2=n^{-1} \sum_{i=1}^n \|\dot{f}_*^{[-i]}-\dot f_i\|^2$ ($\dot{f}$ is the time derivative of $f$);

\item Amplitude mean squared error: $E3=n^{-1} \sum_{i=1}^n d_{a}(q^{[-i]*},q_i)^2$, where $q^{[-i]*},\ q_i$ are the square-root slope transforms of $f^{[-i]*},\ f_i$; 
\item Phase mean squared error: $E4=n^{-1} \sum_{i=1}^n d^{int}_{p}(q^{[-i]*},q_i)^2$. 
\item $L^2$ prediction error: $E5=n^{-1} \sum_{i=1}^n \|f^{[-i]*}-f_i\|^2$.
\end{itemize}
The first three are amplitude errors while the fourth one is the phase error. The last metric is simply based on the standard root mean squared error.

\begin{figure}[!t]
    \centering

    \includegraphics[scale=0.26]{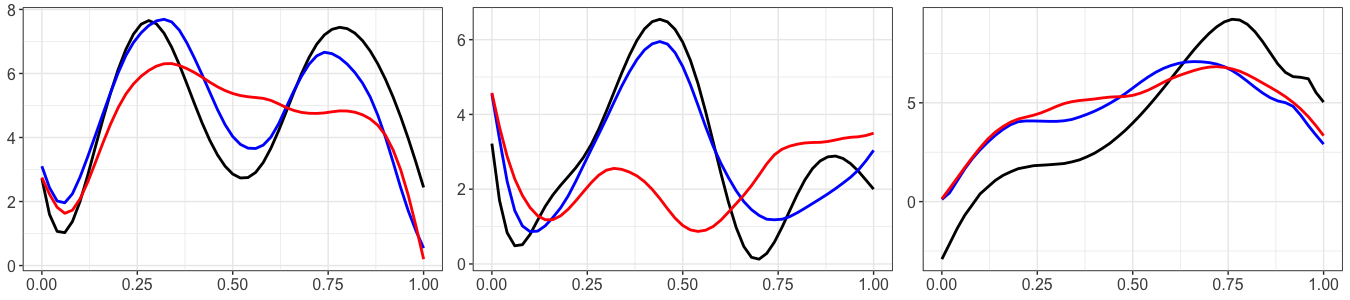}
    \caption{Three example predictions obtained via ordinary kriging (red) and amplitude-phase kriging (blue). The truth is in black.}
    \label{fig:simu_kriging_map2_bsp}
\end{figure}

The advantage of amplitude-phase kriging over ordinary kriging is summarized in Table \ref{tab:kriging_simu}. The improvement in amplitude errors is large when significant phase variation is present in the data. In general, ordinary kriging fails to capture important features of functions in the predictions in presence of phase variation, e.g., the ordinary kriging predictions in Figure \ref{fig:simu_kriging_map2_bsp} do not capture the valley, peak or inflection points in the true function, and tend to result in `flat' predictions. The amplitude-phase kriging, on the other hand, successfully captures these features as well as their magnitude. This results in significant decreases of the various amplitude and phase error metrics. The B-spline data exhibits much more shape variation in the generated functions. This is the more challenging setting for our method. Nonetheless, the proposed approach still outperforms ordinary kriging in most cases, even when phase variation is small. 

While the proposed approach does not outperform ordinary kriging in terms of the $L^2$ prediction error (and Amplitude least squares for the B-spline data), it has been noted in \citet{srivastava2016functional} that the $L^2$ distance, which is used to define these two error metrics, is not a good measure of amplitude and/or phase differences. Furthermore, since ordinary kriging is optimal under the $L^2$ metric, the results based on these measures are naturally biased toward this method. We note that although ordinary kriging has smaller $L^2$ prediction errors for the results shown in Figure \ref{fig:simu_kriging_map2_bsp}, it is clear that the generated predictions are not satisfactory. Appendix \ref{sec:s3.1} contains additional illustrations of the amplitude-phase predictions as well as more detailed results.

\begin{table}[!t]
  \centering
  \caption{Average prediction errors using metrics $E1$-$E5$, across 50 different replicates, for (a) amplitude-phase kriging and (b) ordinary kriging. $E4$ is multiplied by 100 to adjust the scale.}   \footnotesize
    \begin{tabular}{cccccccccccc}
    \hline
    \hline
          & B     & \multicolumn{2}{c}{$E1$}   & \multicolumn{2}{c}{$E2$}  & \multicolumn{2}{c}{$E3$}  & \multicolumn{2}{c}{$E4$}  & \multicolumn{2}{c}{$E5$} \\
          &&(a)&(b)&(a)&(b)&(a)&(b)&(a)&(b)&(a)&(b)\\ \hline
    B-spline & 0.5 & 1.53 & 1.19 & 221 & 225 & 2.32 & 2.44  & 9.95 & 9.69 & 2.60 & 1.83 \\
          & 1  & 1.64 & 1.53 & 311 & 355 & 2.44 & 3.14  & 12.20 & 14.80 & 3.53 & 2.64\\
    Bimodal & 0.5 & 1.08 & 2.46 & 45 & 90 & 0.56 & 1.03 & 1.51 & 1.76 & 11.00 & 9.38 \\
          & 1  & 1.34 & 7.25 & 152 & 400  & 0.84  & 3.99 & 4.58 & 7.65 & 25.40 & 17.70 \\
          \hline
          \hline
    \end{tabular}%
  \label{tab:kriging_simu}%
\end{table}%

\subsection{Clustering}
We next assess the proposed spatial clustering approach on simulated data. 
Let $n$ denote the number of spatial sites where data was observed and $I$ the number of clusters. Then, $n=\sum_{i=1}^I n_i$, where $n_i$ is the number of functions in cluster $i$. Motivated by the fact that amplitude and phase in real data scenarios may exhibit different clustering patterns, we simulate the true clustering of observations with respect to amplitude and phase separately. Our aim is to validate that the proposed amplitude-phase clustering method is able to reveal the true underlying partitions of both amplitude and phase simultaneously, irrespective of whether the spatial partitions of each component agree with each other.

We consider two different designs: (1) where amplitude and phase cluster partitions are the same (agree), and (2) where they are not (disagree). In the first design, the simulated sites are on a $4\times 4$ grid, and are partitioned into four equally sized clusters via the horizontal and vertical lines; here, the amplitude and phase components of the data have the same spatial partition. In the second design, 30 sites are chosen uniformly on $[0,4]^2$. The amplitudes are partitioned by the lines $x=2$ and $y=2$, while the phases are partitioned by the lines $y=x$ and $y=4-x$; see Appendix \ref{sec:s3.2} for a pictorial description of the two designs.


Let $f_{ij}$ be the $j$th functional observation in cluster $i$. We generate spatial functional data with domain $[0,1]$ as $f_{ij}=(a_{ij}\mu+e_{ij})\circ \gamma_{ij}$ ($i=1,...,I; j=1,...,n_i$). We set $\mu(t)=-\cos(2\pi t)$, $a_{ij}=i\delta_a+\epsilon_{a,ij}$, and $\gamma_i$ as the distribution function of $Beta(1,e^{b_{ij}})$, where $b_{ij}=i\delta_b+\epsilon_{b,ij}$; $\delta_a$ and $\delta_b$ are fixed parameters that control the amplitude and phase differences between clusters, respectively. The vector $\{\epsilon_{a,ij}\}$ is generated from a multivariate normal distribution with a mean vector $(5,...,5)^\T$ and Matern covariance $C_{Mat}(\cdot,\cdot; \sigma_a^2,0.5,\ell)$. The vector $\{\epsilon_{b,ij}\}$ follows the correlated uniform distribution on $[-B,B]^n$ with the same correlation range $\ell$; $e_{ij}$ is a zero mean Gaussian process with a diagonal covariance. We fix $\sigma_a^2=1$, $B=1$, $\sigma_e=0.5$ and $\ell=2\surd 2$, and repeat each clustering simulation 100 times. We compare the proposed approach to the standard $L^2$ distance-based method \citep{giraldo2012hierarchical}. The means and standard deviations of rand indices \citep{rand1971objective} for each design, and different choices of $\delta_a$ and $\delta_b$, are shown in Table \ref{tab:clust_simu}. 

The proposed approach outperforms the $L^2$ distance-based method in all scenarios, even when the amplitude and phase partitions agree. When the true clusterings are different, the amplitude-phase clustering is far superior, especially for the larger value of $\delta_a$. The $L^2$ approach is always forced to compromise between the true amplitude and phase clusters, while the proposed approach treats them separately. The $L^2$ metric is sensitive to phase differences. As a result, when $\delta_b$ is large, it captures the phase clustering and exhibits similar performance to the proposed method in that regard. However, it is unable to recover the true amplitude clusters.  

\begin{table}[!t]
  \centering
  \caption{Average rand indices for estimated partitions based on (a) separate amplitude-phase clustering and (b) $L^2$ clustering, with respect to the true amplitude and phase clusters, for two designs, with standard deviations in parentheses.}
  \footnotesize
    \begin{tabular}{ccccccc}
    \hline
    \hline
          &       &       & \multicolumn{2}{c}{Agree} & \multicolumn{2}{c}{Disagree} \\
          $\delta_a$ & $\delta_b$ &Method & Amplitude & Phase & Amplitude & Phase \\\hline
    1     & 0.1   & (a)   & 0.821 (0.100) & 0.769 (0.090) & 0.793 (0.101) & 0.763 (0.076) \\
          &       & (b)    & 0.756 (0.086) & 0.756 (0.086) & 0.719 (0.074) & 0.712 (0.069) \\
          & 0.5   & (a)   & 0.868 (0.088) & 0.961 (0.052) & 0.771 (0.092) & 0.888 (0.078) \\
          &       & (b)    & 0.885 (0.080) & 0.885 (0.080) & 0.705 (0.053) & 0.866 (0.078) \\
    2     & 0.1   & (a)   & 0.944 (0.070) & 0.757 (0.086) & 0.917 (0.071) & 0.737 (0.069) \\
          &       & (b)    & 0.805 (0.083) & 0.805 (0.083) & 0.754 (0.073) & 0.730 (0.068) \\
          & 0.5   & (a)   & 0.945 (0.074) & 0.944 (0.065) & 0.828 (0.085) & 0.907 (0.074) \\
          &       & (b)    & 0.919 (0.074) & 0.919 (0.074) & 0.716 (0.051) & 0.889 (0.073) \\
    \hline
    \hline
    \end{tabular}%
  \label{tab:clust_simu}%
\end{table}%

\section{Real data analysis}

\subsection{Kriging of daily ozone data in north California}

We apply the proposed amplitude-phase kriging method to the U.S. daily ozone data, available on the air data website (https://www.epa.gov/outdoor-air-quality-data) of the United States Environmental Protection Agency. We focus on an area in North California ($35^{\circ}\sim 39^{\circ}$ N, $120\sim 123^{\circ}$ W) with $24$ observation stations. Each station recorded daily average ozone concentration (parts per million) for the year 2018. 
We first smooth the data using smoothing splines. We estimate a Matern variogram based on the trace-variogram, with scale and range estimated by ordinary least squares, and the smoothing parameter and nugget fixed to $0.5$ and $0$, respectively.

To compare the predictive performance of amplitude-phase kriging to that of ordinary kriging, we use leave-one-out cross-validiation on the $24$ observations. We report the mean of the five error metrics, $E1$-$E5$, for each approach in Table \ref{ozone_kriging_1e04}. 
The proposed method outperforms ordinary kriging in terms of all of the reported error metrics. The amplitude-phase kriging amplitude and phase mean squared errors are reduced by $16\%$ and $8\%$, respectively, compared to ordinary kriging. Surprisingly, the proposed method outperforms ordinary kriging in terms of the $L^2$ prediction error, which is the criterion that ordinary kriging optimizes. This shows that combining separate amplitude and phase predictions has a clear advantage in real data scenarios. Appendix \ref{sec:s4.1} contains additional results and illustrations.

This kriging analysis was performed on spatial functional data located in a small region, on which the isotropic assumption is realistic. The proposed prediction method can be generalized to universal kriging on a large area following ideas of \citep{caballero2013universal,menafoglio2013universal} who consider non-stationary functional random fields. 

\begin{table}[!t] 
\centering 
  \caption{Average leave-one-out cross-validation prediction errors using (a) amplitude-phase kriging and (b) ordinary kriging for ozone data in North California. All numbers were multiplied by $1000$.} 
  \label{ozone_kriging_1e04}
\begin{tabular}{cccccccccc} 
\hline
\hline
  \multicolumn{2}{c}{$E1$} & \multicolumn{2}{c}{$E2$} & \multicolumn{2}{c}{$E3$} & \multicolumn{2}{c}{$E4$} & \multicolumn{2}{c}{$E5$}\\
  (a)&(b)&(a)&(b)&(a)&(b)&(a)&(b)&(a)&(b)\\\hline
 4.71 & 4.83 & 1.59e-03 & 1.83e-03 & 3.32 & 3.98 & 70.26 & 76.69 & 6.64 & 6.67 \\ 
\hline
\hline
\end{tabular} 
\end{table}

\subsection{Clustering of Canadian weather data}

Next, we apply the proposed amplitude-phase clustering method to the Canadian weather data \citep{ramsay2004functional}. The data can be found in the R package 'fda' \citep{ramsay2020fdapackage}. In this paper, we analyze the daily temperature data averaged over 1960-1994, collected at 35 stations in Canada. Because the 35 stations cover a large area, the stationarity assumption here is likely violated, e.g., compared to longitude, latitude generally has a larger effect on the temperature due to its relationship to the duration and angle of solar radiation. Thus, before modelling the spatial dependency in this dataset, we first filter out the longitudinal and latitudinal trends to make the data approximately stationary on the entire spatial domain. To do this, we fit a functional linear regression model where longitude and latitude are included as covariates; the same approach was taken in \citet{giraldo2012hierarchical}. The resulting functional residuals are then smoothed using smoothing splines and used as the data for clustering. 

\begin{figure}
    \centering
    \includegraphics[scale=0.35]{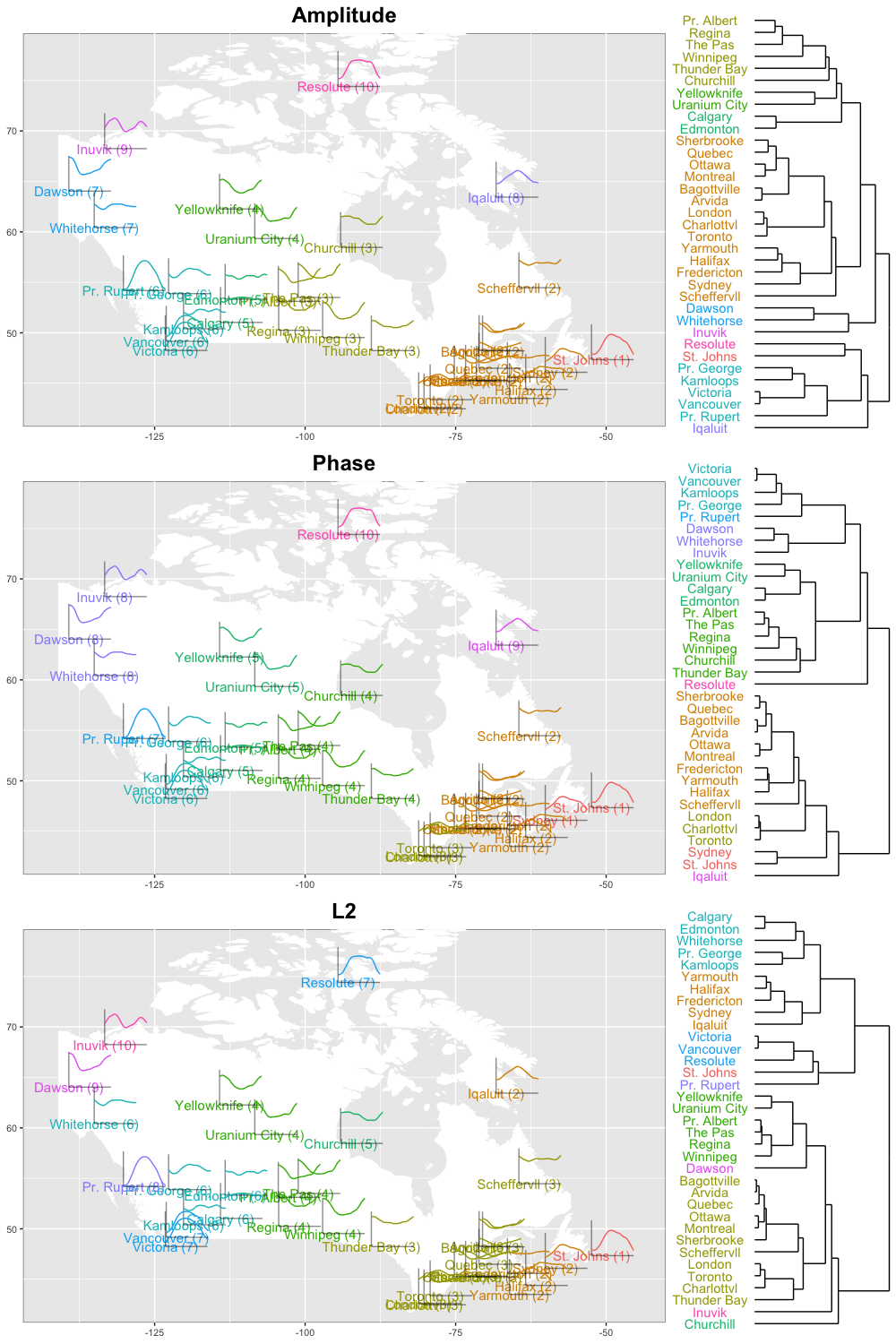}
    \caption{Clustering (average linkage, 10 clusters in different colors) of functional residuals, after adjusting for latitude and longitude effects, obtained from the Canadian weather data.}
    \label{fig:CA_average}
\end{figure}

We use the clustering method described in Section \ref{sec:cluster} and compare the results to the $L^2$ metric-based clustering of \citet{giraldo2012hierarchical}. As in the previous section, we estimate a Matern variogram for all methods. 
The hierarchical clustering trees as well as the clustering results on the map of Canada are shown in Figure \ref{fig:CA_average}. Based on separate clustering of amplitude and phase, we discover some interesting results. First, the amplitude and phase clusterings agree in the middle of Canada and have local differences in the West and Southeast regions. Second, in the Northwest, the phase clustering groups Inuvik, Dawson and Whitehouse together whereas amplitude clustering separates Inuvik from the other two. The shape of the functional residual at Inuvik is different from the other two, which is captured by the amplitude clustering; the phase clustering does not distinguish them because it focuses on the timing of the biggest valley. Third, in the Southeast, the amplitude clustering groups most of the sites together due to the small magnitude of the functional residuals; in contrast, the phase clustering provides a finer partition of this region that is related to the distance of each site from the Atlantic. 
In the $L^2$ clustering, we observe some unnatural results. For example, Resolute, a station in the arctic circle, is clustered with the Vancouver and Victoria stations on the West coast. Also, compared to amplitude-phase clustering, the $L^2$ method generates more single-element clusters due to confounding of amplitude and phase. We also implemented hierarchical clustering without spatial weighting (see Appendix \ref{sec:s4.2} for results). It is clear that involving spatial dependency in the clustering helps preserve connectivity of adjacent sites, making the results more interpretable.

\section{Discussion}
It is difficult to verify the key assumptions of stationarity and isotropy for spatial functional data, especially when one decouples amplitude and phase components, which effectively results in two sets of functional data. Despite this, when deviation from stationarity is not too large, the amplitude and phase trace-variograms provide useful summary statistics of spatial variation. Results from simulations and real data analyses offer corroboration. 

Extensions of developments in this paper to the setting of noisy, sparse spatial functional data constitute ongoing work. Results here represent the first foray towards analyzing spatial complex functional data objects such as shapes of curves \citep{srivastava2016functional} and surfaces \citep{JKLS} by decoupling spatial, shape and nuisance variations.

\section{Acknowledgement}
Funding through multiple grants from the National Science Foundation and a grant from the National Cancer Institute at the National Institutes of Health is gratefully acknowledged. 



\bibliography{spatial_ref.bib}

\begin{thebibliography}{33}
\expandafter\ifx\csname natexlab\endcsname\relax\def\natexlab#1{#1}\fi

\bibitem[{Abramowicz et~al.(2017)Abramowicz, Arnqvist, Secchi, De~Luna, Vantini
  \& Vitelli}]{abramowicz2017clustering}
\textsc{Abramowicz, K.}, \textsc{Arnqvist, P.}, \textsc{Secchi, P.},
  \textsc{De~Luna, S.~S.}, \textsc{Vantini, S.} \& \textsc{Vitelli, V.} (2017).
\newblock Clustering misaligned dependent curves applied to varved lake
  sediment for climate reconstruction.
\newblock \textit{Stochastic environmental research and risk assessment}
  \textbf{31}, 71--85.

\bibitem[{Caballero et~al.(2013)Caballero, Giraldo \&
  Mateu}]{caballero2013universal}
\textsc{Caballero, W.}, \textsc{Giraldo, R.} \& \textsc{Mateu, J.} (2013).
\newblock A universal kriging approach for spatial functional data.
\newblock \textit{Stochastic environmental research and risk assessment}
  \textbf{27}, 1553--1563.

\bibitem[{Chakraborty \& Panaretos(2020)}]{CP}
\textsc{Chakraborty, A.} \& \textsc{Panaretos, V.} (2020).
\newblock Functional registration and local variations: Identifiability, rank,
  and tuning.
\newblock \textit{Bernoulli (to appear)} .

\bibitem[{Cressie \& Wikle(2011)}]{cressie2011statistics}
\textsc{Cressie, N.} \& \textsc{Wikle, C.~K.} (2011).
\newblock \textit{Statistics for Spatio-Temporal Data}.
\newblock John Wiley \& Sons.

\bibitem[{Delicado et~al.(2010)Delicado, Giraldo, Comas \&
  Mateu}]{delicado2010statistics}
\textsc{Delicado, P.}, \textsc{Giraldo, R.}, \textsc{Comas, C.} \&
  \textsc{Mateu, J.} (2010).
\newblock Statistics for spatial functional data: some recent contributions.
\newblock \textit{Environmetrics: The official journal of the International
  Environmetrics Society} \textbf{21}, 224--239.

\bibitem[{Everitt \& Landau(2001)}]{everitt2001m}
\textsc{Everitt, B.} \& \textsc{Landau, S.~L.} (2001).
\newblock M. 2001. cluster analysis.
\newblock \textit{Arnold, London} .

\bibitem[{Giraldo et~al.(2011)Giraldo, Delicado \& Mateu}]{giraldo2011ordinary}
\textsc{Giraldo, R.}, \textsc{Delicado, P.} \& \textsc{Mateu, J.} (2011).
\newblock Ordinary kriging for function-valued spatial data.
\newblock \textit{Environmental and Ecological Statistics} \textbf{18},
  411--426.

\bibitem[{Giraldo et~al.(2012)Giraldo, Delicado \&
  Mateu}]{giraldo2012hierarchical}
\textsc{Giraldo, R.}, \textsc{Delicado, P.} \& \textsc{Mateu, J.} (2012).
\newblock Hierarchical clustering of spatially correlated functional data.
\newblock \textit{Statistica Neerlandica} \textbf{66}, 403--421.

\bibitem[{Goulard \& Voltz(1993)}]{goulard1993geostatistical}
\textsc{Goulard, M.} \& \textsc{Voltz, M.} (1993).
\newblock Geostatistical interpolation of curves: a case study in soil science.
\newblock In \textit{Geostatistics Tr{\'o}ia?92}. Springer, pp. 805--816.

\bibitem[{Haggarty et~al.(2015)Haggarty, Miller \&
  Scott}]{haggarty2015spatially}
\textsc{Haggarty, R.}, \textsc{Miller, C.} \& \textsc{Scott, E.} (2015).
\newblock Spatially weighted functional clustering of river network data.
\newblock \textit{Journal of the Royal Statistical Society: Series C (Applied
  Statistics)} \textbf{64}, 491--506.

\bibitem[{Ibragimov \& Rozanov(2012)}]{ibragimov2012gaussian}
\textsc{Ibragimov, I.~A.} \& \textsc{Rozanov, Y.~A.} (2012).
\newblock \textit{Gaussian random processes}, vol.~9.
\newblock Springer Science \& Business Media.

\bibitem[{Jermyn et~al.(2017)Jermyn, S., Laga \& Srivastava}]{JKLS}
\textsc{Jermyn, I.~H.}, \textsc{S., K.}, \textsc{Laga, H.} \&
  \textsc{Srivastava, A.} (2017).
\newblock \textit{Elastic shape analysis of three-dimensional objects}.
\newblock Morgan and Claypool publishers.

\bibitem[{Kurtek \& Srivastava(2011)}]{KS}
\textsc{Kurtek, S.} \& \textsc{Srivastava, A.} (2011).
\newblock Signal estimation under random time-warpings and nonlinear signal
  alignment.
\newblock In \textit{Proceedings of Advances in Neural Information Processing
  Systems, NIPS}. pp. 676--683.

\bibitem[{Marron et~al.(2015)Marron, Ramsay, Sangalli \&
  Srivastava}]{marron2015functional}
\textsc{Marron, J.~S.}, \textsc{Ramsay, J.~O.}, \textsc{Sangalli, L.~M.} \&
  \textsc{Srivastava, A.} (2015).
\newblock Functional data analysis of amplitude and phase variation.
\newblock \textit{Statistical Science} , 468--484.

\bibitem[{Mateu \& Romano(2017)}]{mateu2017advances}
\textsc{Mateu, J.} \& \textsc{Romano, E.} (2017).
\newblock Advances in spatial functional statistics.

\bibitem[{Menafoglio \& Petris(2016)}]{menafoglio2016kriging}
\textsc{Menafoglio, A.} \& \textsc{Petris, G.} (2016).
\newblock Kriging for hilbert-space valued random fields: The operatorial point
  of view.
\newblock \textit{Journal of Multivariate Analysis} \textbf{146}, 84--94.

\bibitem[{Menafoglio et~al.(2018)Menafoglio, Pigoli \&
  Secchi}]{menafoglio2018kriging}
\textsc{Menafoglio, A.}, \textsc{Pigoli, D.} \& \textsc{Secchi, P.} (2018).
\newblock Kriging riemannian data via random domain decompositions.
\newblock \textit{arXiv preprint arXiv:1812.07435} .

\bibitem[{Menafoglio et~al.(2013)Menafoglio, Secchi, Dalla~Rosa
  et~al.}]{menafoglio2013universal}
\textsc{Menafoglio, A.}, \textsc{Secchi, P.}, \textsc{Dalla~Rosa, M.} et~al.
  (2013).
\newblock A universal kriging predictor for spatially dependent functional data
  of a hilbert space.
\newblock \textit{Electronic Journal of Statistics} \textbf{7}, 2209--2240.

\bibitem[{Nerini et~al.(2010)Nerini, Monestiez \&
  Mant{\'e}}]{nerini2010cokriging}
\textsc{Nerini, D.}, \textsc{Monestiez, P.} \& \textsc{Mant{\'e}, C.} (2010).
\newblock Cokriging for spatial functional data.
\newblock \textit{Journal of Multivariate Analysis} \textbf{101}, 409--418.

\bibitem[{Ramsay(2004)}]{ramsay2004functional}
\textsc{Ramsay, J.~O.} (2004).
\newblock Functional data analysis.
\newblock \textit{Encyclopedia of Statistical Sciences} \textbf{4}.

\bibitem[{Ramsay et~al.(2020)Ramsay, Graves \& Hooker}]{ramsay2020fdapackage}
\textsc{Ramsay, J.~O.}, \textsc{Graves, S.} \& \textsc{Hooker, G.} (2020).
\newblock Package ‘fda’ .

\bibitem[{Rand(1971)}]{rand1971objective}
\textsc{Rand, W.~M.} (1971).
\newblock Objective criteria for the evaluation of clustering methods.
\newblock \textit{Journal of the American Statistical association} \textbf{66},
  846--850.

\bibitem[{Reyes et~al.(2015)Reyes, Giraldo \& Mateu}]{reyes2015residual}
\textsc{Reyes, A.}, \textsc{Giraldo, R.} \& \textsc{Mateu, J.} (2015).
\newblock Residual kriging for functional spatial prediction of salinity
  curves.
\newblock \textit{Communications in Statistics-Theory and Methods} \textbf{44},
  798--809.

\bibitem[{Romano et~al.(2010)Romano, Balzanella \&
  Verde}]{romano2010clustering}
\textsc{Romano, E.}, \textsc{Balzanella, A.} \& \textsc{Verde, R.} (2010).
\newblock Clustering spatio-functional data: a model based approach.
\newblock In \textit{Classification as a Tool for Research}. Springer, pp.
  167--175.

\bibitem[{Romano et~al.(2017)Romano, Balzanella \& Verde}]{romano2017spatial}
\textsc{Romano, E.}, \textsc{Balzanella, A.} \& \textsc{Verde, R.} (2017).
\newblock Spatial variability clustering for spatially dependent functional
  data.
\newblock \textit{Statistics and Computing} \textbf{27}, 645--658.

\bibitem[{Sangalli et~al.(2010)Sangalli, Secchi, Vantini \&
  Vitelli}]{sangalli2010k}
\textsc{Sangalli, L.~M.}, \textsc{Secchi, P.}, \textsc{Vantini, S.} \&
  \textsc{Vitelli, V.} (2010).
\newblock K-mean alignment for curve clustering.
\newblock \textit{Computational Statistics \& Data Analysis} \textbf{54},
  1219--1233.

\bibitem[{Schmidt et~al.(2011)Schmidt, Guttorp \&
  O'Hagan}]{schmidt2011considering}
\textsc{Schmidt, A.~M.}, \textsc{Guttorp, P.} \& \textsc{O'Hagan, A.} (2011).
\newblock Considering covariates in the covariance structure of spatial
  processes.
\newblock \textit{Environmetrics} \textbf{22}, 487--500.

\bibitem[{Secchi et~al.(2013)Secchi, Vantini \& Vitelli}]{secchi2013bagging}
\textsc{Secchi, P.}, \textsc{Vantini, S.} \& \textsc{Vitelli, V.} (2013).
\newblock Bagging voronoi classifiers for clustering spatial functional data.
\newblock \textit{International journal of applied earth observation and
  geoinformation} \textbf{22}, 53--64.

\bibitem[{Srivastava \& Klassen(2016)}]{srivastava2016functional}
\textsc{Srivastava, A.} \& \textsc{Klassen, E.~P.} (2016).
\newblock \textit{Functional and shape data analysis}, vol. 475.
\newblock Springer.

\bibitem[{Srivastava et~al.(2011)Srivastava, Wu, Kurtek, Klassen \&
  Marron}]{srivastava2011registration}
\textsc{Srivastava, A.}, \textsc{Wu, W.}, \textsc{Kurtek, S.}, \textsc{Klassen,
  E.} \& \textsc{Marron, J.~S.} (2011).
\newblock Registration of functional data using fisher-rao metric.
\newblock \textit{arXiv preprint arXiv:1103.3817} .

\bibitem[{Stein(1988)}]{stein1988asymptotically}
\textsc{Stein, M.~L.} (1988).
\newblock Asymptotically efficient prediction of a random field with a
  misspecified covariance function.
\newblock \textit{The Annals of Statistics} , 55--63.

\bibitem[{Wang et~al.(2016)Wang, Chiou \& M{\"u}ller}]{wang2016functional}
\textsc{Wang, J.-L.}, \textsc{Chiou, J.-M.} \& \textsc{M{\"u}ller, H.-G.}
  (2016).
\newblock Functional data analysis.
\newblock \textit{Annual Review of Statistics and Its Application} \textbf{3},
  257--295.

\bibitem[{Yakowitz \& Szidarovszky(1985)}]{yakowitz1985comparison}
\textsc{Yakowitz, S.} \& \textsc{Szidarovszky, F.} (1985).
\newblock A comparison of kriging with nonparametric regression methods.
\newblock \textit{Journal of Multivariate Analysis} \textbf{16}, 21--53.

\end{thebibliography}

\clearpage

\begin{appendices}

\section{Proposition Proofs}
\subsection{Proofs of Proposition 1 and 3} \label{apdex:prop13}

We first prove Proposition 1. The proof of Proposition 3 follows along almost identical lines and is omitted. 
Let $\langle \cdot , \cdot \rangle$ denote the $L^2$ inner-product and $\|\cdot\|$ the corresponding ${L}^2$ norm. Then, the prediction error decomposes as follows:
\begin{align}\label{eq:decomp1}
&\E(\|\tilde q_0-(q_0,\gamma_0)\|^2)
=\E\left[\left\|\sum\limits_{i=1}^n \eta_i \{(q_i,\hat\gamma_i)-(q_0,\gamma_0)\}\right\|^2\right]\nonumber\\
=&\E\left[\sum\limits_{i=1}^n \sum\limits_{j=1}^n \eta_i\eta_j\langle (q_i,\hat\gamma_i)-(q_0,\gamma_0), (q_j,\hat\gamma_j)-(q_0,\gamma_0) \rangle\right]\nonumber\\
=&\sum\limits_{i=1}^n \sum\limits_{j=1}^n \eta_i\eta_j\int_0^1\E[ \{(q_i,\hat\gamma_i)(u)-(q_0,\gamma_0)(u)\}\{(q_j,\hat\gamma_j)(u)-(q_0,\gamma_0)(u)\}]du,
\end{align}
where the second equality holds due to the constraint $\sum_{i=1}^n \eta_i=1$, and the third equality uses Fubini's theorem. Under the assumptions made in the main article, all aligned functions have a common expectation, say $q$, and a common variance function. Thus, for any $i,j=1,...,n$,

\begin{align} \label{eq:decomp2}
& \int_0^1\E [\{(q_i,\hat\gamma_i)(u)-(q_0,\gamma_0)(u)\}\{(q_j,\hat\gamma_j)(u)-(q_0,\gamma_0)(u)\}]du\nonumber\\
=&\int_0^1\E [\{(q_i,\hat\gamma_i)(u)-q(u)+q(u)-(q_0,\gamma_0)(u)\}\nonumber\\
&\ \ \ \ \ \  \{(q_j,\hat\gamma_j)(u)-q(u)+q(u)-(q_0,\gamma_0)(u)\}]du\nonumber\\
=&\int_0^1\cov\{(q_i,\hat\gamma_i)(u),(q_j,\hat\gamma_j)(u)\}-\cov\{(q_i,\hat\gamma_i)(u),(q_0,\gamma_0)(u)\} \nonumber\\
&\ \ \ \ -\cov\{(q_0,\gamma_0)(u),(q_j,\hat\gamma_j)(u)\} +\var\{(q_0,\gamma_0)(u)\}du\nonumber\\
=& \int_0^1 \var\{(q_0,\gamma_0)(u)\}- \cov\{(q_0,\gamma_0)(u),(q_j,\hat\gamma_j)(u)\}du\nonumber\nonumber\\
& +\int_0^1 \var\{(q_0,\gamma_0)(u)\}- \cov\{(q_0,\gamma_0)(u),(q_i,\hat\gamma_i)(u)\}du\nonumber\\
& -\int_0^1 \var\{(q_0,\gamma_0)(u)\}- \cov\{(q_i,\hat\gamma_i)(u),(q_j,\hat\gamma_j)(u)\}du\nonumber\\
=&V_a(h_{0j})+V_a(h_{i0})-V_a(h_{ij}),
\end{align} 
where $h_{ij}=\|s_i-s_j\|$, for $i,j=1,...,n$. The last equality results from the definition of the amplitude trace-variogram, which by Fubini's theorem and the assumption that $\var\{(q_s,\gamma_s)(t)\}=\sigma^2(t)$ does not depend on the spatial location $s\in \mathcal D$ implies
\begin{align*}
&V_a(h_{ij})=\frac12\E(\|(q_i,\hat\gamma_i)-(q_j,\hat\gamma_j)\|^2)\\
=&\int_0^1 \var\{(q_0,\gamma_0)(u)\}- \cov\{(q_i,\hat\gamma_i)(u),(q_j,\hat\gamma_j)(u)\} du.  
\end{align*} 
Plugging (\ref{eq:decomp2}) into (\ref{eq:decomp1}) results in $\eta^\T\mathcal V_a \eta$, where $\mathcal V_a=[V_a(h_{0j})+V_a(h_{i0})-V_a(h_{ij})]_{n\times n}$.


\subsection{Proof of Proposition 2}
Without loss of generality, let $s_1$ be the closest site to $s_0$. Then, the template is chosen as $q_1(t)=(c_1\mu_q,\gamma_1^{-1})(t)$. We align each $q_i$ to $q_1$ using

\begin{align*}
\hat \gamma_i=&\underset{\gamma}\argmin\ \|(q_i,\gamma)-q_1\|^2=
\underset{\gamma}\argmin\  \{\|(q_i,\gamma)\|^2-2\langle(q_i,\gamma),q_1\rangle  + \|q_1\|^2\}\\
=&\underset{\gamma}\argmin\  \{\|q_i\|^2-2\langle(q_i,\gamma),q_1 \rangle + \|q_1\|^2\} =  \underset{\gamma}\argmax\ 
\langle(q_i,\gamma),q_1 \rangle\\
=&\underset{\gamma}\argmax\  c_1c_i\langle(\mu_q,\gamma_i^{-1}\circ\gamma),(\mu_q,\gamma_1^{-1}) \rangle = \gamma_i\circ\gamma_1^{-1}
\end{align*}
The third equality is due to the norm-preserving action of $\Gamma$ on $\mathcal Q$. The aligned functions are then given by $(q_i,\hat \gamma_i)(t)=c_i (\mu_q,\gamma_1^{-1})(t)$ for $i=0,1,...,n$. Note that $q_0$ is unknown, and thus we do not know the aligned function $(q_0,\hat \gamma_0)(t)$. But, we know that $(q_0,\hat \gamma_0)\in [q_0]$ regardless of what $\hat \gamma_0$ is. Thus, it is sufficient to show that $\tilde q_0\to(q_0,\hat \gamma_0)$ in $L^2$ as $n\to \infty$. The coefficient $ \eta$ is estimated by minimizing the amplitude prediction error
\begin{align*}
&\E \left(\|\sum\limits_{i=1}^n\eta_i (q_i,\hat \gamma_i) - (q_0,\hat \gamma_0)\|^2\right)
=\E\left(\|(c_0-\sum\limits_{i=1}^n\eta_ic_i) (\mu_q,\gamma_1^{-1})\|^2\right)\\
&=\E\left\{(c_0-\sum\limits_{i=1}^n\eta_ic_i)^2\|(\mu_q,\gamma_1^{-1})\|^2\right\}\\
&=\E\left\{(c_0-\sum\limits_{i=1}^n\eta_ic_i)^2\|\mu_q\|^2 \right\}\\
&=\E \left\{(c_0-\sum\limits_{i=1}^n\eta_ic_i)^2\right\} \|\mu_q\|^2.
\end{align*}
The last equality again uses the fact that the action of $\Gamma$ on $\mathcal Q$ is norm-preserving. After eliminating the phase variation, amplitude kriging is equivalent to univariate kriging of the scaling coefficients.
In this way, the consistency of $\hat c_0=\sum_{i=1}^n\hat \eta_ic_i$ determines the consistency of amplitude kriging.
When the variogram is known and continuous in a neighborhood of 0, and a limit point of $\{  s_1,...,  s_n\}$ is $s_0$ as $n\to \infty$, the expected prediction error $\E\{(\hat c_0-c_0)^2\}\to0$ as $n \to\infty$ \citep{yakowitz1985comparison}. 

The condition of knowing the variogram of $\{c_i\}$ can be relaxed. \citet{yakowitz1985comparison} and \citet{stein1988asymptotically} discuss the effects of misspecification of the variogram on kriging. Given the true covariance function $C$, the best linear unbiased estimator is $\hat c_0$. If the covariance function is misspecified as $C^*$, we have the best estimator $\hat c^*_0$. \citet{stein1988asymptotically} demonstrate that if $C^*$ and $C$ are compatible \citep{ibragimov2012gaussian}, then 
\begin{equation*}
\frac{\E\{(\hat c_0-c_0)^2\}}{\E\{(\hat c^*_0-c_0)^2\}}\to 1, \quad \frac{\E\{(\hat c_0-c_0)^2\}}{\E^*\{(\hat c^*_0-c_0)^2\}}\to 1, \quad  n\to\infty,
\end{equation*}
where $E^*(\cdot)$ is the expectation under the misspecified covariance $C^*$. Since the prediction error $\E\{(\hat c_0-c_0)^2\}\to 0$ as $n\to \infty$ given the true variogram of $\{c_i\}$, the prediction error when using $C^*$ also converges to 0, if $C^*$ is compatible with $C$.

\section{Implementation Details}
\subsection{Regularized alignment} \label{sec:s2.1}
In this section, we provide further details of the implementation of our approach for the kriging simulation and real data studies. When we estimate the phase component of each function for phase kriging, we utilize a regularized alignment method \citep{srivastava2016functional} to increase robustness to noise or large shape variation in the observed spatial functional data. The optimal warping function that aligns $q_2$ to $q_1$ is estimated using:
\begin{equation}\label{eq:penalizedalign}
\gamma^*=\underset{\gamma\in\Gamma}\argmin \ \{ \|q_1-(q_2,\gamma)\|^2+\lambda \|\psi-\psi_{id}\|^2 \}, 
\end{equation}
where $\psi={\dot\gamma}^{1/2}$ is the squared root slope transformed $\gamma$ and $\psi_{id}(t)=1$ is the squared root slope transformed identity warping $\gamma_{id}(t)=t$. The tuning parameter $\lambda$ is selected in each run of leave-one-out cross-validation by additional five-fold cross-validation. Specifically, the leave-one-out cross-validation training set is randomly divided into five folds. Given a candidate $\lambda$, we use each fold as the prediction target, and use the data in the other four folds to realize the phase kriging prediction. We compute the sum of squared extrinsic phase distances (see Definition 2 in the main article) between the predicted and true phase functions. We repeat the five-fold cross-validation ten times and select the optimal $\lambda$ as the one that minimizes the average prediction error. Despite the complex hierarchical cross-validation struction of phase kriging, the computation is very efficient in practice. 

\subsection{Enlarged space for phase variogram}\label{sec:s2.2}

\begin{figure}
    \centering
    \includegraphics[scale=0.15]{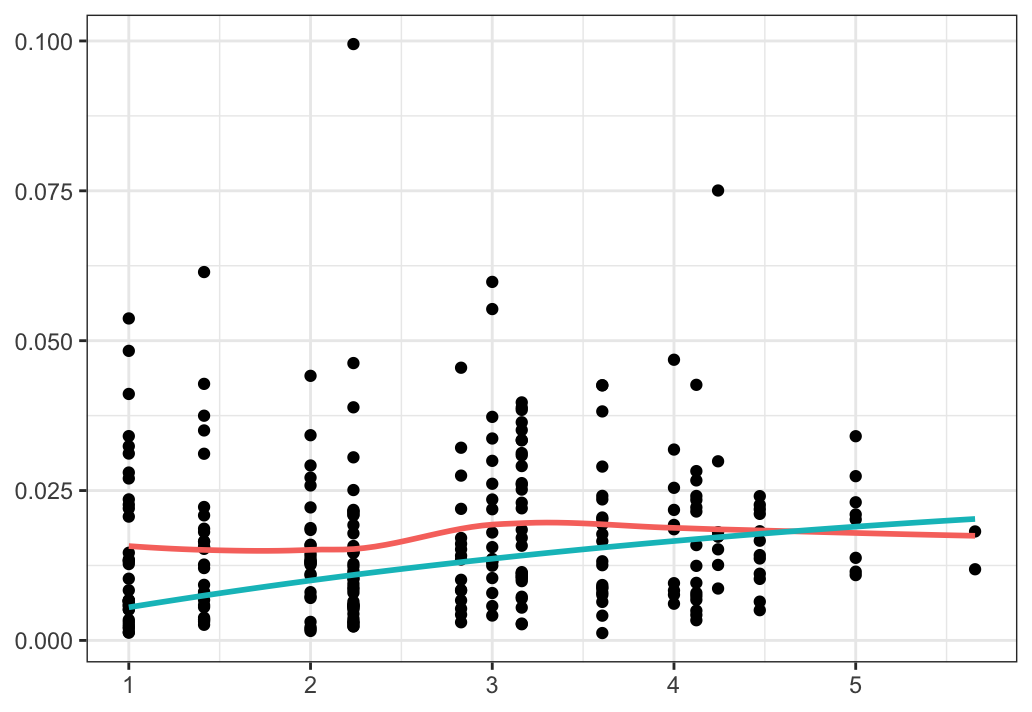}
    \includegraphics[scale=0.15]{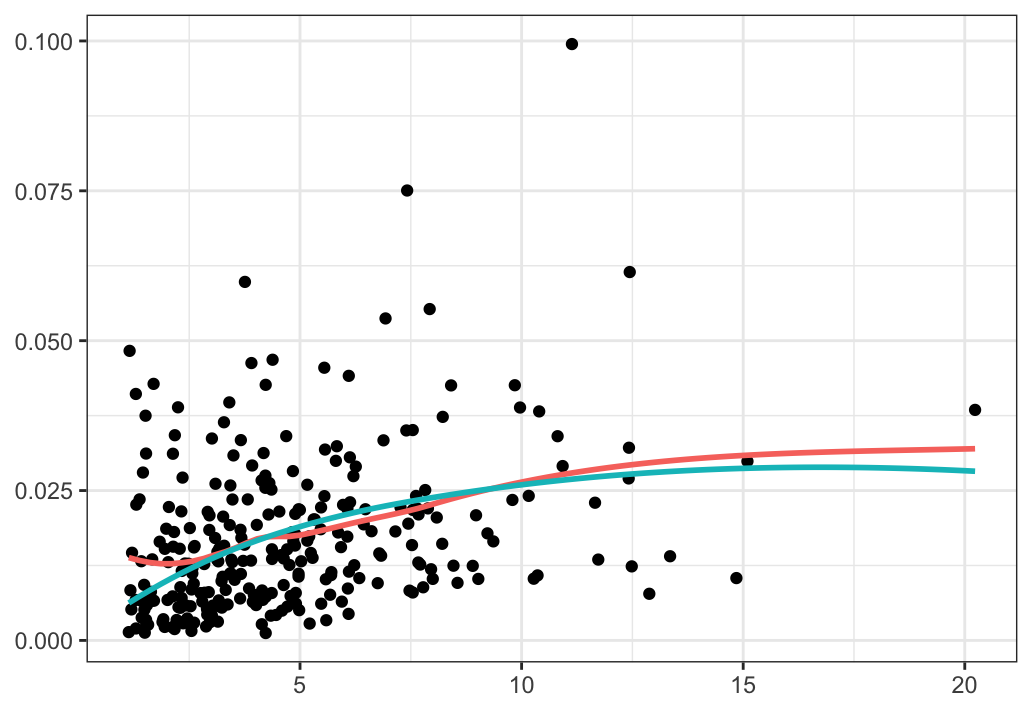}
    \caption{Estimated phase trace-variograms for simulated data, with spatially correlated phase components, using the spatial domain only (left) and the enlarged domain with $\omega=503.2$ (right). The green line is the fitted Mat\'ern variogram while the red line is the empirical variogram.}
    \label{fig:vargm_elarge}
\end{figure}

The tuning parameter $\omega$, in the definition of the enlarged space distance in (4) in the main article, takes the value that maximizes the goodness of fit of the parametric variogram. Additionally, if the least squares-based fitting of the variogram results in a smaller reduction of the error than $5\%$, due to the introduction of the enlarged space, we set $\omega=0$; in this case, we only use the spatial coordinates to define the lag. Based on empirical experiments, when the phase component contains negligible spatial dependence using the spatial lag only, but is correlated with the shape of the observed functions, the estimated $\omega$ can be fairly large. In this case, the phase trace-variogram is consistent with the dependence present in the shapes of the observed functions. Figure \ref{fig:vargm_elarge} presents a comparison of two fitted phase trace-variograms for simulated spatial functional data: the left panel uses the spatial lag only while the right panel uses the enlarged domain with function shape information. The true phase components in the simulated data have strong spatial correlation and are independent of the amplitude errors. The improvement in goodness-of-fit in the estimated variogram via introducing the enlarged space is clear.  

\section{Additional Details for Simulation Studies}

\subsection{Kriging}\label{sec:s3.1}

\begin{figure}[!t]
\centering
\includegraphics[scale=0.2]{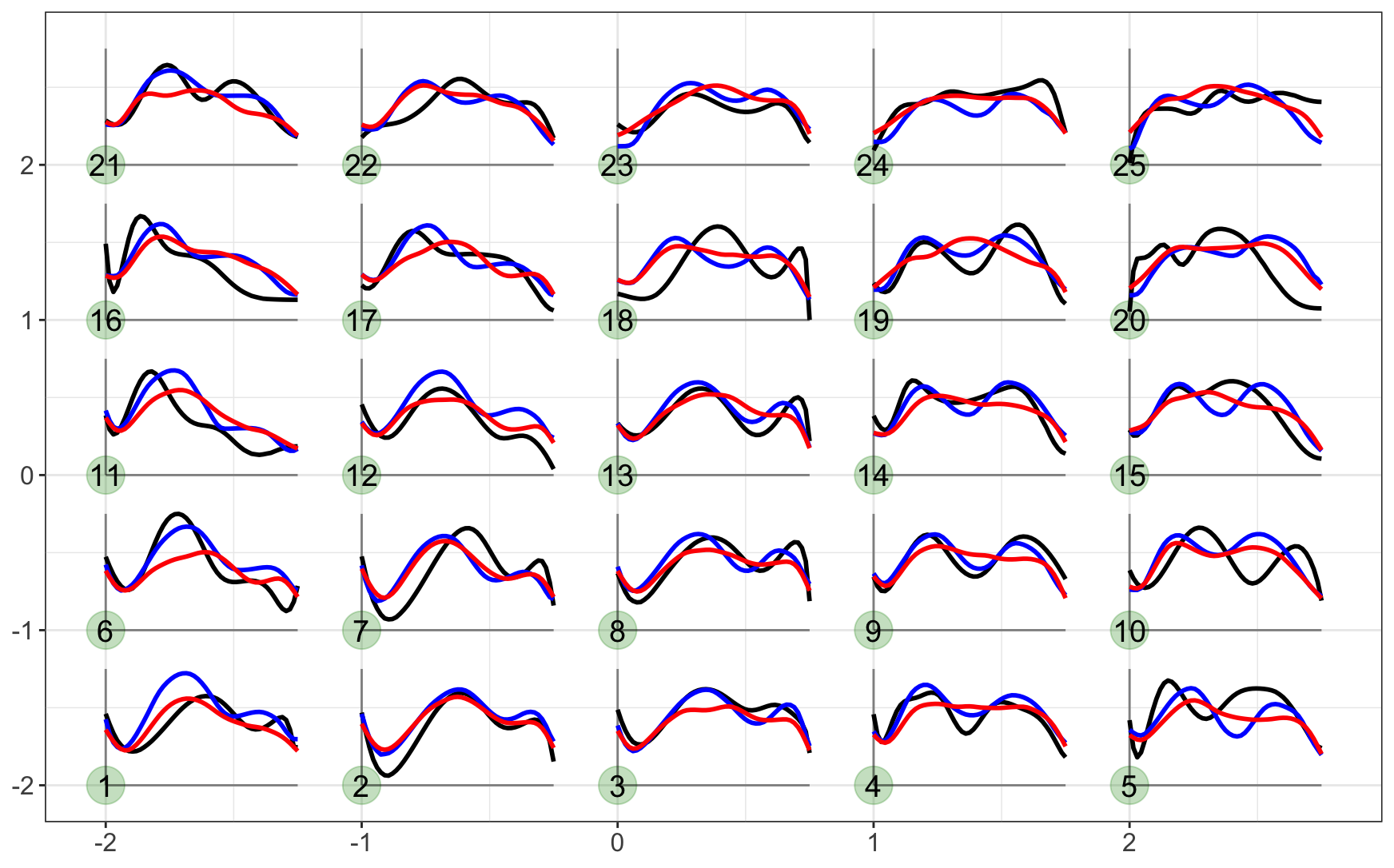}
\caption{Leave-one-out cross-validation results for one simulation run with amount of phase variation set to $B=2$. The average error metrics $E3$, $E4$ and $E5$, with respect to the truth (black), in this run are 1.86, 0.067, 3.914 for amplitude-phase kriging (blue) and 3.56, 0.119, 3.5 for ordinary kriging, respectively. The numbered green points on the map indicate the simulated spatial locations.}
\label{fig:kriging_simu}
\end{figure}

In the main article, we show several typical examples of kriging to reveal why amplitude-phase kriging outperforms ordinary kriging. In Figure \ref{fig:kriging_simu}, we display the complete leave-one-out kriging results on the simulated B-spline data. Compared to ordinary kriging, amplitude-phase kriging shows a clear advantage in estimating the shape of functions. For example, at sites 8, 9, 13 and 14, ordinary kriging fails to estimate the valley in the true functions. On the other hand, the proposed method provides a much better estimate. Despite the poor performance of ordinary kriging in these examples, the prediction error $E5$, which is based on the $L^2$ metric, results in a smaller error for ordinary kriging. This is mainly due to the inappropriateness of the $L^2$ metric for measuring differences between functions in the presence of phase variation; it tends to place a very small penalty on ``flat'' functions. The error of amplitude-phase kriging mainly results from the prediction of phase, which is especially challenging when predicting the functions at the boundary of the spatial domain since they have fewer neighbors.

To assess the effects of the distribution of sites on the spatial domain on prediction results, we perform the same simulation using spatial functional data located at 25 randomly (uniformly) sampled sites on the domain $[-2,2]^2$. The data generating process is the same as in Section 6.1 in the main article. As in the main article, we consider different scales for the phase variation, which is controlled by the parameter $B$. Additionally, we evaluate our approach across different spatial range parameters $\ell$ in the amplitude and phase spatial covariances (set to be equal across the two components). The prediction results are reported in Table \ref{tab:kriging_simu_scat}. This simulation leads to similar conclusions as those that were reached based on the equally-spaced simulation settings. Some additional findings are as follows. The prediction errors are smaller when sites are randomly scattered on the spatial domain, because some of the sites tend to have many nearby neighbors, increasing prediction accuracy. Further, strong spatial dependency enhances prediction performance of both methods, as expected, but does not change the relative performance of the two methods.

\begin{table}[!t]
  \centering
  \caption{Average prediction errors using metrics $E1$-$E5$, across 50 different replicates when sites are randomly sampled on the spatial domain, for (a) amplitude-phase kriging and (b) ordinary kriging. $E4$ is multiplied by 100 to adjust the scale.}
  \scriptsize
    \begin{tabular}{cccccccccccccc}
    \hline
    \hline
          & B  & $\ell$   & \multicolumn{2}{c}{$E1$}   & \multicolumn{2}{c}{$E2$}  & \multicolumn{2}{c}{$E3$}  & \multicolumn{2}{c}{$E4$}  & \multicolumn{2}{c}{$E5$} \\
          &&&(a)&(b)&(a)&(b)&(a)&(b)&(a)&(b)&(a)&(b)\\\hline
    \multicolumn{1}{l}{B-spline} & 0.5   & $2\surd 2$ & 1.18  & 0.88  & 237   & 233   & 1.87  & 2.04  & 8.49  & 8.24  & 2.15  & 1.43 \\
          &       & $3\surd 2$ & 0.86  & 0.62  & 212   & 205   & 1.73  & 1.71  & 6.70   & 6.12  & 1.64  & 0.99 \\
          & 1     & $2\surd 2$ & 1.20   & 1.11  & 239   & 274   & 2.03  & 2.46  & 9.75  & 11.10  & 2.81  & 2.13 \\
          &       & $3\surd 2$ & 0.92  & 0.86  & 228   & 242   & 1.84  & 2.12  & 9.40   & 9.84  & 2.32  & 1.63 \\
    \multicolumn{1}{l}{Bimodal} & 0.5   & $2\surd 2$ & 1.00     & 1.44  & 36  & 56  & 0.45  & 0.69  & 1.26  & 1.41  & 7.00     & 5.34 \\
          &       & $3\surd 2$ & 0.78  & 1.04  & 32  & 44  & 0.41  & 0.53  & 1.11  & 1.17  & 6.66  & 4.32 \\
          & 1     & $2\surd 2$ & 1.17  & 5.38  & 215   & 383   & 0.81  & 2.90   & 4.01  & 5.47  & 21.00    & 13.80 \\
          &       & $3\surd 2$ & 1.02  & 4.70   & 194   & 350   & 0.74  & 2.47  & 3.76  & 4.73  & 23.00    & 14.00 \\\hline\hline
    \end{tabular}%
  \label{tab:kriging_simu_scat}%
\end{table}%
\color{black}

\subsection{Clustering}\label{sec:s3.2}
\begin{figure}[!t]
    \centering
    \begin{subfigure}{0.4\textwidth}
      \centering
        \includegraphics[scale=.38]{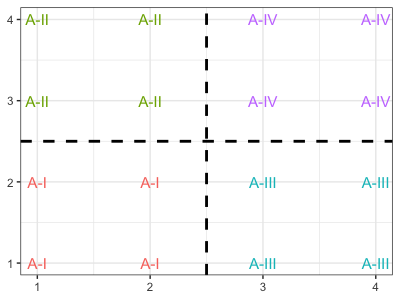}
      \caption{}
    \end{subfigure}%
    \begin{subfigure}{0.4\textwidth}
      \centering
         \includegraphics[scale=.38]{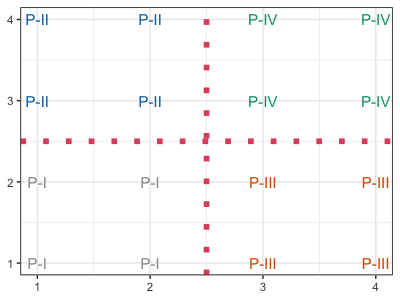}
      \caption{}
    \end{subfigure}
        \begin{subfigure}{0.4\textwidth}
      \centering
         \includegraphics[scale=.38]{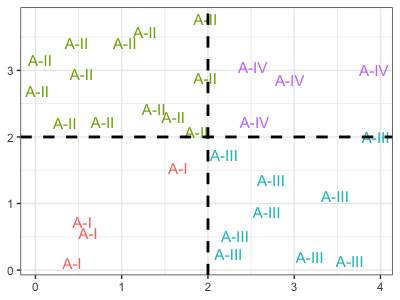}
      \caption{}
    \end{subfigure}
    \begin{subfigure}{0.4\textwidth}
      \centering
         \includegraphics[scale=.38]{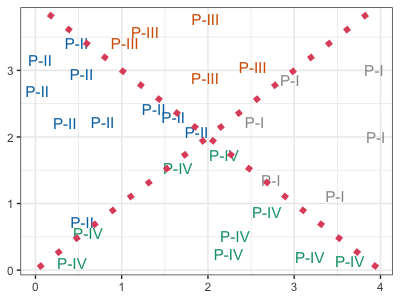}
      \caption{}
    \end{subfigure}
    \caption{The ``agree'' design for equally-spaced spatial sites: (a) amplitude partition and (b) phase partition. The ``disagree'' for randomly (uniformly) sampled spatial sites: (a) amplitude partition and (b) phase partition. The black dashed lines are the boundaries of the different amplitude clusters and the red dotted lines are the boundaries of the phase clusters.}
    \label{fig:truecluster}    
\end{figure}

\begin{figure}[!t]
    \centering
    \begin{subfigure}{0.3\textwidth}
      \centering
        \includegraphics[scale=.37]{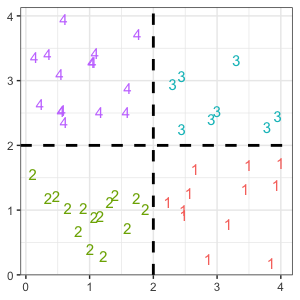}
      \caption{}
    \end{subfigure}%
    \begin{subfigure}{0.3\textwidth}
      \centering
         \includegraphics[scale=.37]{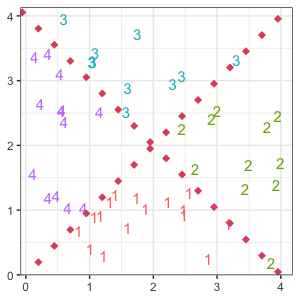}
      \caption{}
    \end{subfigure}
        \begin{subfigure}{0.3\textwidth}
      \centering
         \includegraphics[scale=.37]{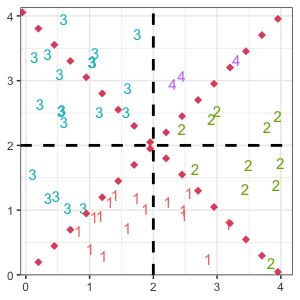}
      \caption{}
    \end{subfigure}
    \caption{Estimated partitions using amplitude-phase clustering for (a) amplitude and (b) phase. (c) Estimated partition using $L^2$ clustering. These results correspond to a single simulated run for the ``disagree'' design. The black dashed lines are the boundaries of the true amplitude clusters and the red dotted lines are the boundaries of the true phase clusters.}
    \label{fig:estcluster}    
\end{figure}

In the clustering simulation in the main article, we use two designs for forming the true partitions, which we refer to as ``agree'' (amplitude and phase partitions are the same) and ``disagree'' (amplitude and phase partitions are different). Figure \ref{fig:truecluster} shows the two designs pictorially with the partitions highlighted by different colors. Figure \ref{fig:estcluster} shows one example of estimated clusters obtained based on amplitude-phase clustering and $L^2$ clustering, when the true clusters in amplitude and phase are different. Again, the estimated clusters are highlighted in different colors. In the presence of amplitude and phase variation, $L^2$ clustering is always dominated by one of the components, resulting in a mixture of the true amplitude and phase partitions. On the other hand, amplitude-phase clustering decouples these two sources of variability and is able to estimate the true amplitude and phase clusterings simultaneously.

\section{Additional results for real data kriging and clustering}

\subsection{Kriging of North California Ozone Data} \label{sec:s4.1}
Kriging is a local interpolation method that essentially uses a weighted mean of neighboring observations as a prediction. The spatial dependency directly determines the contribution of each observation in the prediction under the second-order stationary and isotropic assumptions. In our method, this procedure is separate for amplitude and phase. We show in detail how the amplitude-phase kriging procedure works on the ozone data example from the main article.

\begin{figure}[!t]
\begin{center}
\includegraphics[width=0.9\textwidth]{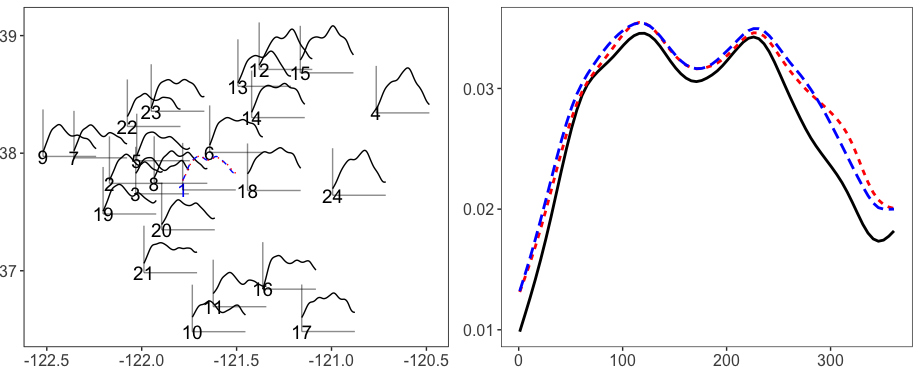}
\caption{Prediction of average daily ozone concentration in 2018 at site 1 (Livermore, CA) via leave-one-out cross-validation. Left: Prediction map with solid lines indicating observed functions and the dashed line corresponding to the predictions at site 1. Right: Zoom-in of the site 1 predictions; the true function is given in black, the amplitude-phase prediction in blue and the ordinary kriging prediction in red. The amplitude, phase, and $L^2$ prediction errors ($E3$/$E4$/$E5$) for this single predicted function are 0.0345/0.115/0.00179 based on amplitude-phase kriging and  0.046/0.151/0.00195 based on ordinary kriging.}
\label{fig:krigingsite1}
\end{center}
\end{figure}

\begin{figure}[!t]
\begin{center}
\includegraphics[scale=0.16]{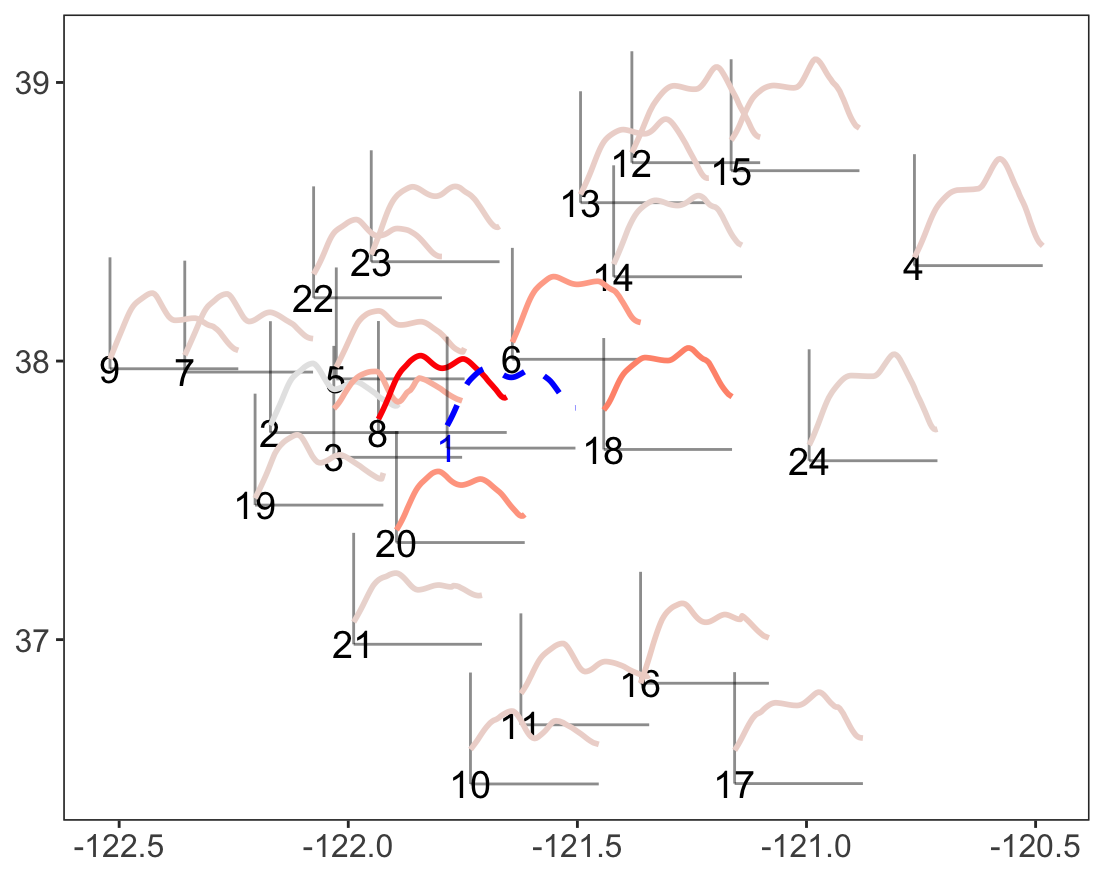}
\includegraphics[scale=0.16]{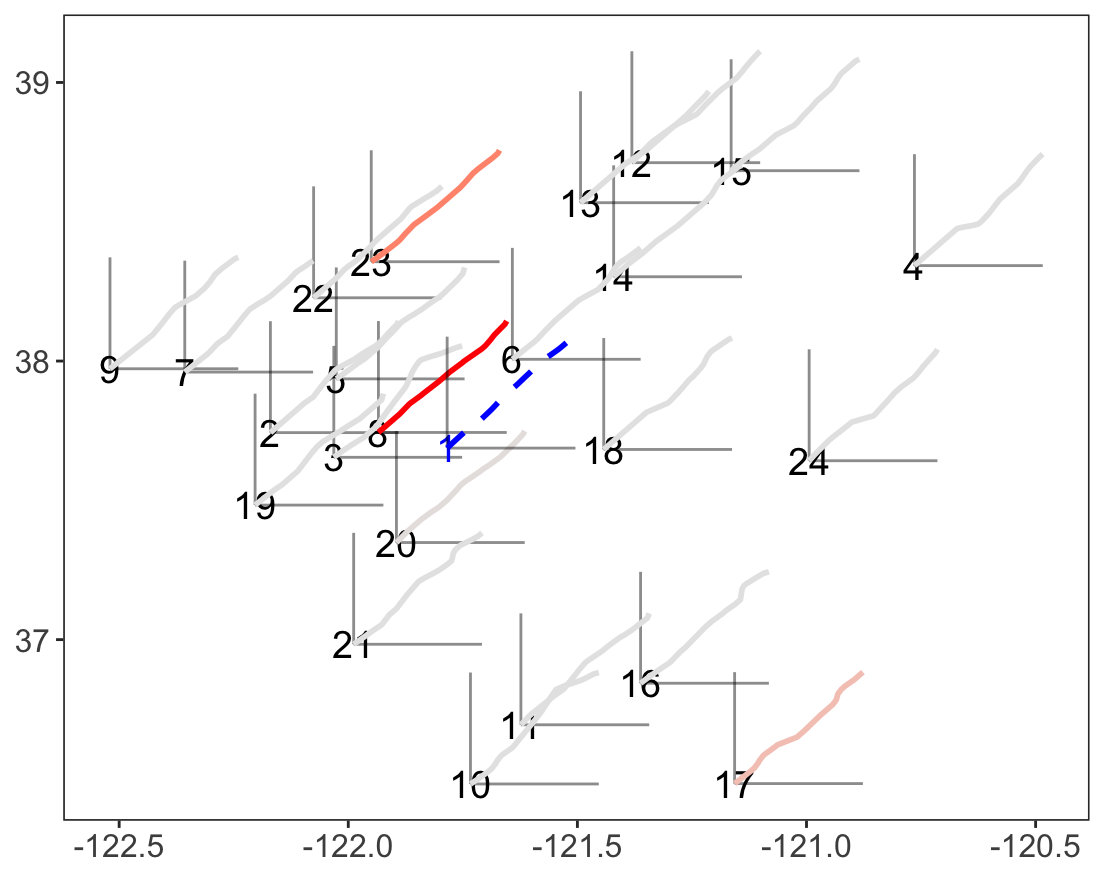}
\caption{Kriging of amplitude (left) and phase  (right) components at site 1 (Livermore, CA): the solid curves are the amplitude and phase components at observed site; dash line are the predicted amplitude and phase components. The depth of color shows the contribution of component (from 0 to 1) at each site.}
\label{fig:apkmap}
\end{center}
\end{figure}

In Figure \ref{fig:krigingsite1}, we show the result of a single cross-validation run to predict the function at site 1 in Livermore, CA ($37.68^{\circ}$ N, $121.77^{\circ}$W). We also show the observed functions at their relative locations on the map of California; each of the 24 spatial sites is labeled by a number. In Figure \ref{fig:apkmap}, we show the  the kriging maps for the amplitude and phase components separately. The shading of the plotted functions on each map corresponds to the contribution (weight) of each function in the final kriging estimate. Amplitude kriging generally borrows information from neighboring sites since we only use the spatial coordinates (distance) to model the dependency in this case. On the other hand, in phase kriging, we use the enlarged space, which includes the spatial locations and the shape of the observed functions, to model the dependency. Thus, the highest contribution into the final kriging estimate is a combination of phase functions that are nearby and those that correspond to observed functions that have a similar shape to the predicted amplitude. The phase variation in this dataset is small, and is mainly due to local delays or advances in the timeline, which represent small fadeviations from identity warping. Predicting the phase component is a difficult task in practice since its definition depends on the shape of functional data. Furthermore, the signal/spatial dependency in the phase component is generally fairly weak. This is why many previous studies prefer to treat phase variability as noise. However, in this real data analysis, we have found that even if the phase signal is not as strong as the amplitude signal, separate amplitude and phase prediction is still beneficial as evidenced by the errors reported in the main article. 

Prior to kriging, all functional data is smoothed using a smoothing spline with a pre-specified tuning parameter, $\iota$. The cross-validation results using different values for this tuning parameter are reported in the Table \ref{tab:kriging_ozone_sm}. When the input data is relatively smooth ($\iota=5\times 10^{-4}$), the difference in amplitude errors between the proposed method and ordinary kriging is very small. This is due to the smoothing procedure, which smooths out notable shape characteristics of the functions. Nonetheless, the proposed method still has smaller amplitude prediction errors on average. In contrast, when less smoothing is applied to the observed functions ($\iota=1\times 10^{-4}$), the amplitude prediction errors are much smaller for the proposed method as compared to ordinary kriging; this is because more shape characteristics of the observed functions are preserved. In terms of phase, the gains due to our method are very large for all values of the smoothing parameter. 
Overall, the performance of amplitude-phase kriging under different degrees of smoothing is fairly robust. 

\begin{table}[!t] 
\centering 
  \caption{Leave-one-out cross-validation average prediction errors for (a) amplitude-phase kriging and (b) ordinary kriging, for the ozone data in North California. All numbers were multiplied by $1000$.} 
  \label{tab:kriging_ozone_sm}
\footnotesize  
\begin{tabular}{ccccccccccc}
\hline
\hline
  &\multicolumn{2}{c}{$E1$} & \multicolumn{2}{c}{$E2$} & \multicolumn{2}{c}{$E3$} & \multicolumn{2}{c}{$E4$} & \multicolumn{2}{c}{$E5$}\\
  $\iota$ &(a)&(b)&(a)&(b)&(a)&(b)&(a)&(b)&(a)&(b)\\\hline
 0.001 &4.71 & 4.83 & 1.59e-03 & 1.83e-03 & 3.32 & 3.98 & 70.26 & 76.69 & 6.64 & 6.67 \\ 
  0.003 &4.17 & 4.98 & 7.41e-04 & 8.01e-04 & 1.96 & 2.23 & 46.80 & 53.64 & 7.08 & 6.44 \\
   0.005 &4.02 & 4.09 & 6.82e-04 & 6.43e-04 & 1.62 & 1.62 & 43.00 & 51.21 & 6.18 & 6.32 \\ \hline \hline
\end{tabular} 
\end{table}

\begin{figure}[!t]
    \centering
    \includegraphics[scale=0.34]{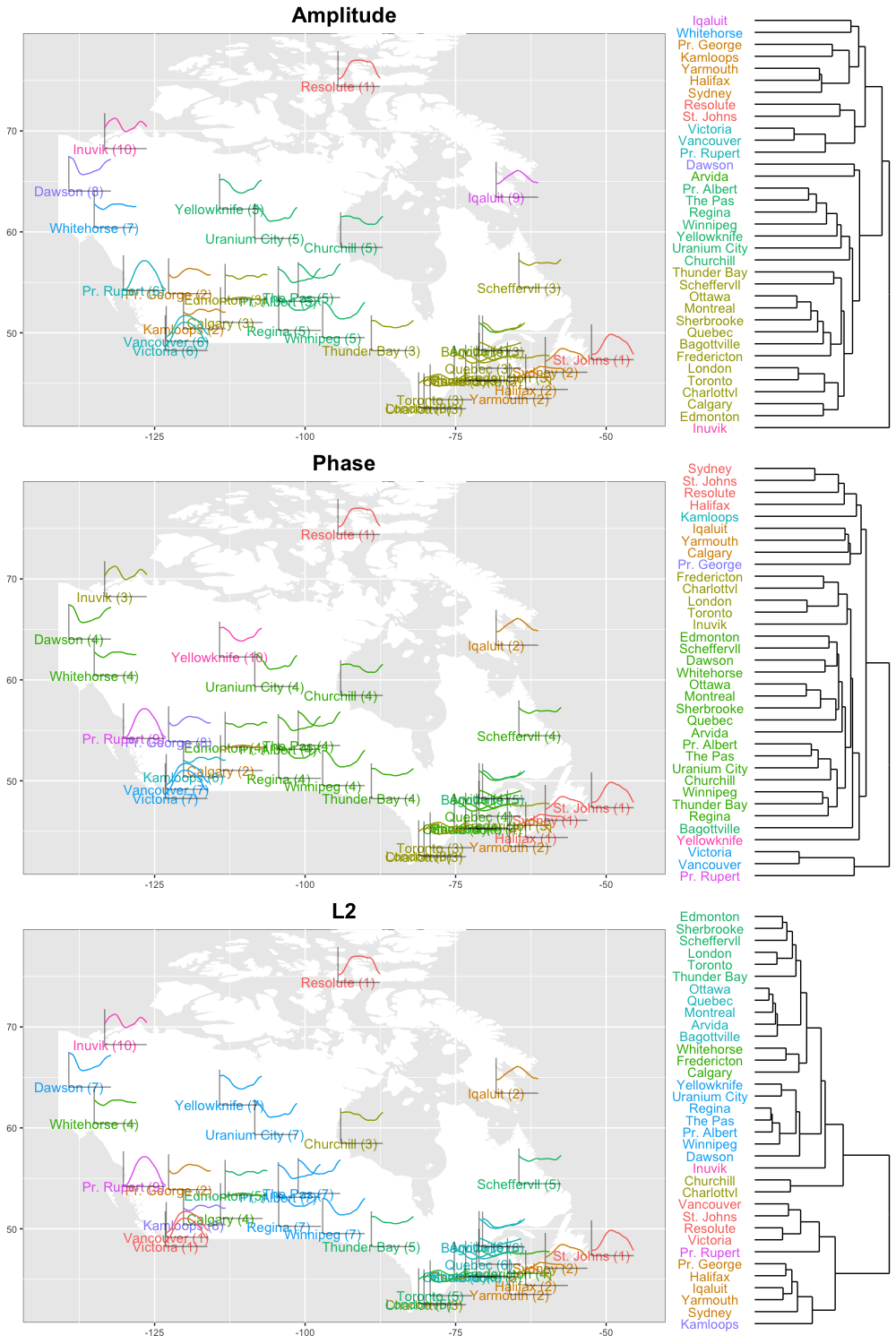}
    \caption{Clustering (average linkage, ten clusters in different colors) of functional residuals, after adjusting for latitude and longitude effects, obtained from the Canadian weather data without adjusting for spatial dependency.}
    \label{fig:nonspatial}
\end{figure}

\subsection{Clustering of Canadian weather data without adjusting for spatial dependency}\label{sec:s4.2}
The clustering techniques discussed in the main article account for the spatial dependency across functional observations. The spatial dependency is encoded in the dissimilarity matrix via weights, where the discrepancy between pairs of functions near each other, with respect to the distance on the domain, is down-weighted. The weight, computed using the fitted trace-variogram, works as an empirical prior, where the dependency decays as the domain distance increases, and its range and rate of decay depend on the data. In particular, if the data suggest no spatial dependency, spatially weighted clustering is the same as standard hierarchical clustering without spatial information. 
Spatial weighting tends to preserve connectivity of clusters and often results in more interpretable results.

To show the difference between spatially-weighted clustering and standard clustering, we applied standard hierarchical clustering, with average linkage, to the same data as considered in Section 7.2 in the main article. The results are shown in Figure \ref{fig:nonspatial}, and are directly comparable to Figure 4 in the main article. The partitions estimated using hierarchical clustering without accounting for spatial dependency tend to be more scattered, e.g., the green cluster in phase clustering. Such results are difficult to interpret and relate to geological factors. In addition, standard hierarchical clustering of this data results in many more single-element clusters. 

\end{appendices}

\end{document}